\documentclass[final,pageno]{jpaper}
\pdfoutput=1
\usepackage[normalem]{ulem}
\usepackage[T1]{fontenc}
\usepackage{libertine}
\usepackage{pifont}
\usepackage{inconsolata}

\usepackage{mystyle}
\usepackage{macros}
\usepackage{algpseudocode}
\usepackage{pifont}
\usepackage{color}

\usepackage{setspace}
\begin{document}

\title{\Large{A Framework for Accelerating Bottlenecks in GPU Execution\\with Assist Warps}}
\author{Nandita Vijaykumar\quad Gennady Pekhimenko\quad Adwait Jog$^\dag$\quad
Saugata Ghose\quad Abhishek
Bhowmick\quad \\
Rachata Ausavarungnirun\quad Chita Das$^\dag$\quad Mahmut Kandemir$^\dag$\quad
Todd C. Mowry\quad Onur Mutlu
\vspace{1mm}\\
%\vspace{0.5cm}
\textbf{Carnegie Mellon University} \quad\quad\quad\quad\quad\textbf{$^\dag$ Pennsylvania State
University} 
\vspace{1mm}\\
  %	Pittsburgh, PA \quad\quad\quad\quad\quad\quad\quad\quad\quad$^\dag$ University Park, PA \\
	\authemail{\normalsize{\{nandita,ghose,abhowmick,rachata,onur\}@cmu.edu}}    \\
	 \authemail{\normalsize{\{gpekhime,tcm\}@cs.cmu.edu}} \quad\quad
\authemail{\normalsize{\{adwait,das,kandemir\}@cse.psu.edu}}
}

% commented out
\comm{\author{Nandita Vijaykumar\quad Gennady Pekhimenko\quad Adwait Jog$^\dag$\quad Abhishek
Bhowmick\quad \\
Rachata Ausavarungnirun\quad Onur Mutlu\quad Chita Das$^\dag$\quad Mahmut Kandemir$^\dag$\quad
Todd C. Mowry\\\\
  \textbf{Carnegie Mellon University}\quad\quad$^\dag$\textbf{Pennsylvania State
University}\\ } } 
\comm{
 \begin{minipage}{8cm}   \centering
    \textbf{Carnegie Mellon University}\\
    \authemail{\{nandita,abhowmick,rachata,onur\}@cmu.edu}\\
    \authemail{\{gpekhime,tcm\}@cs.cmu.edu}\\
  \end{minipage}
  \begin{minipage}{6cm}
    \centering
    \textbf{Pennsylvania State University}\\
    \authemail{\{adwait,das,kandemir\}@cse.psu.edu}\\
  \end{minipage}
}
\comm{
\author{
  \begin{minipage}{4.5cm}
    \centering
    Nandita Vijaykumar\\
    \small{\authemail{nandita@.cmu.edu}}
  \end{minipage}
  \begin{minipage}{4.5cm}
    \centering
    Gennady Pekhimenko\\
    \small{\authemail{gpekhime@cs.cmu.edu}}
  \end{minipage}
  \begin{minipage}{4.0cm}
    \centering
     Adwait Jog$^\dag$\\
    \small{\authemail{adwait@cse.psu.edu}}
  \end{minipage}
  \begin{minipage}{4.5cm}
    \centering
     Abhishek Bhowmick\\
    \small{\authemail{abhowmick@cmu.edu}}
  \end{minipage}\\\\
  \begin{minipage}{4.5cm}
    \centering
     Rachata Ausavarungnirun\\
    \small{\authemail{rachata@cmu.edu}}
  \end{minipage}
  \begin{minipage}{3cm}
    \centering
    Onur Mutlu\\
    \small{\authemail{onur@cmu.edu}}
  \end{minipage}
  \begin{minipage}{3.5cm}
    \centering
     Chita Das$^\dag$\\
    \small{\authemail{das@cse.psu.edu}}
  \end{minipage}
  \begin{minipage}{4.5cm}
    \centering
     Mahmut Kandemir$^\dag$\\
    \small{\authemail{kandemir@cse.psu.edu}}
  \end{minipage}
  \begin{minipage}{4cm}
    \centering
    Todd C. Mowry\\
    \small{\authemail{tcm@cs.cmu.edu}}
  \end{minipage}\\\\ 
  \textbf{Carnegie Mellon University}\quad\quad$^\dag$\textbf{Pennsylvania
State University} 
%\author{
%  Nandita Vijaykumar\quad Gennad\quad Third Author\quad Fourth Author\\
%  Your Affiliation\\
%  \authemail{\{author1,author2,author3,author4\}@affiliation.edu}
%}
}
}

\date{}
\maketitle
%%%%%%%%%%%%%%%%%%%%%%%%%%%%%%%%%%%%%%%%%%%%%%%%%%%%%%%%%%%%%%%%%%%%% 
% The ACM Copyright Paragraph must appear on the first page of each 
% paper. Government authors should refer to the alternative copyright
% instructions @ http://www.acm.org/sigs/volunteer_resources/conference_manual/6-5proc
%%%%%%%%%%%%%%%%%%%%%%%%%%%%%%%%%%%%%%%%%%%%%%%%%%%%%%%%%%%%%%%%%%%%% 
 
%\comm{\begin{figure}[!b]
%  \vspace{3pt} 
%    \footnotesize{ Permission to make digital or hard
%    copies of all or part of this work for personal or classroom
%    use is granted without fee provided that copies are not made
%    or distributed for profit or commercial advantage and that
%    copies bear this notice and the full citation on the first
%    page. Copyrights for components of this work owned by others
%    than ACM must be honored. Abstracting with credit is
%    permitted. To copy otherwise, or republish, to post on servers
%    or to redistribute to lists, requires prior specific
%    permission and/or a fee. Request permissions from
%    \texttt{Permissions@acm.org}.\\
%    ISCA '15, June 13 - 17, 2015, Portland, OR, USA\\
%    \copyright 2015 ACM. ISBN 978-1-4503-3402-0/15/06 \$15.00\\
%    DOI: \texttt{http://dx.doi.org/10.1145/2749469.2750399} }
%\end{figure}
%}

\thispagestyle{empty}

\begin{abstract}
\vspace{0.2cm}
Modern Graphics Processing Units (GPUs) are well provisioned to
support the concurrent execution of thousands of
threads. Unfortunately, different bottlenecks during execution and
heterogeneous application requirements create imbalances in
utilization of resources in the cores. For example, when a GPU is
bottlenecked by the available off-chip memory bandwidth, its computational
resources are often overwhelmingly idle, waiting for data from memory
to arrive.

This work describes the \SADAfull (\SADA) framework that employs
idle on-chip resources to alleviate different bottlenecks in GPU
execution. \SADA provides flexible mechanisms to automatically
generate ``assist warps'' that execute on GPU cores to perform
specific tasks that can improve GPU performance and efficiency.

\SADA enables the use of idle computational units and pipelines to
alleviate the memory bandwidth bottleneck, e.g., by using assist warps
to perform data compression to transfer less data from
memory. Conversely, the same framework can be employed to handle cases
where the GPU is bottlenecked by the available computational units, in
which case the memory pipelines are idle and can be used by \SADA to
speed up computation, e.g., by performing memoization using assist
warps.

We provide a comprehensive design and evaluation \green{of \SADA}
to perform effective and flexible data compression in the GPU memory hierarchy to
alleviate the memory bandwidth bottleneck.
%We show that \SADA can
%flexibly implement multiple different compression algorithms
%(beneficial for different applications) instead of requiring a
%dedicated hardware implementation for each algorithm.  
Our extensive
evaluations show that \SADA, when used to implement data compression,
provides an average performance improvement of 41.7\% (as high as
2.6X) across a variety of memory-bandwidth-sensitive GPGPU
applications.

% mitigating the memory
%bandwidth bottleneck via efficient and flexible data compression on
%the memory bus. We also show that CABA can flexibly implement
%different algorithms and is, therefore, more resource efficient than
%dedicated hardware-based mechanisms. 

We believe that CABA is a flexible framework that enables the use of idle
resources to improve application performance with different optimizations and perform other useful tasks. We
discuss how CABA can be used, for example, for memoization, prefetching,
handling interrupts, profiling, redundant multithreading, and speculative
precomputation.

\end{abstract}

%END
\comm{Modern Graphics Processing Units (GPUs) are well provisioned to support the concurrent execution of hundreds or even thousands of
threads. Unfortunately, different bottlenecks during execution and heterogeneous application requirements create imbalances in utilization of resources in the cores. For example, when a GPU is bottlenecked by the available off-chip
bandwidth, its computational resources are overwhelmingly
idle, waiting for data from memory to arrive. 

This paper makes a case for the \SADAfull (\SADA) framework that employs idle on-chip resources to alleviate
different bottlenecks in GPU execution. \SADA provides flexible mechanisms to
automatically generate ``assist warps'' that execute on GPU cores to perform
specific tasks that can improve GPU performance and efficiency. 

\comm{We provide a
detailed study of how \SADA can be used to perform efficient and flexible data
compression in GPUs to reduce the memory bandwidth bottleneck. }\SADA enables
the use of idle computational units and pipelines to solve the memory
bandwidth problem, e.g., by using assist warps to perform efficient data
compression to transfer less data from memory. Conversely, the same framework
can be extended to cases where the GPU is bottlenecked by the available
computational units, in which case the memory pipelines and resources are idle and can be used to speed up GPU execution, e.g., by
performing prefetching or memoization using assist warps.

  We demonstrate the usefulness of \SADA in mitigating the memory bandwidth
bottleneck via efficient and flexible data compression on the memory bus. We also show that CABA can flexibly implement different algorithms and is, therefore, more resource efficient than dedicated hardware-based mechanisms. Our
extensive evaluations show that \SADA provides an average performance
improvement of 40.6\% (as high as 2.6X) across a variety of
memory-bandwidth-sensitive GPGPU applications.}

\section{Introduction}
\label{sec:intro}

Modern Graphics Processing Units (GPUs) play an important role in
delivering high performance and energy efficiency for many classes of
applications and different computational platforms. GPUs employ
fine-grained multi-threading to hide the high memory access latencies
with thousands of concurrently running threads~\cite{keckler}.  GPUs
are well provisioned with different resources (e.g., SIMD-like
computational units, large register files) to support the execution of
a large number of these hardware contexts. \green{Ideally, if the demand
for all types of resources is properly balanced, all these
resources should be fully utilized by the application}\comm{Ideally, all these
resources should be fully utilized by the application if the demand
for all types of resources is properly balanced}. Unfortunately, this
balance is very difficult to achieve in practice.

As a result, bottlenecks in program execution, e.g., limitations in
memory or computational bandwidth, lead to long stalls and idle
periods in the shader pipelines of modern
GPUs~\cite{veynu,osp-isca13,owl-asplos13,equalizer}.  \green{Alleviating
these bottlenecks with optimizations implemented in
dedicated hardware requires significant engineering cost and effort.
Fortunately, the resulting under-utilization of on-chip computational
and memory resources from these imbalances in application
requirements, offers some new opportunities. For
example, we can use these resources for efficient integration of
\emph{hardware-generated threads} that perform useful work to
accelerate the execution of the primary threads.} Similar \emph{helper
  threading} ideas have been proposed in the context of
general-purpose
processors~\cite{ssmt,ssmt2,assisted-execution,ht4,ddmt,ht20,assisted-execution-04}
to either extend the pipeline with more contexts or use spare hardware
contexts to pre-compute useful information that aids main code
execution (e.g., to aid branch prediction, prefetching, etc.).
%%% ONUR: SMT is too specific. Use hardware contexts.

%%% ONUR: Writing is sloppy in places. ``availability of ...'' did not make
%sense. It was a broken sentence %% ONUR: SIMT is not defined.  %% ONUR:
%``preclude'' was not the right word to use...
We believe that the general idea of helper threading can lead to even
more powerful optimizations and new opportunities in the context of
modern GPUs than in CPUs because (1) the abundance of on-chip
resources in a GPU obviates the need for idle hardware
contexts~\cite{ht5,ht4} or the addition of more storage (registers,
rename tables, etc.) and compute units~\cite{slice,ssmt} required to
handle more contexts and (2) the relative simplicity of the GPU
pipeline avoids the complexities of handling register renaming,
speculative execution, precise interrupts, etc.~\cite{ssmt2}. However,
\green{GPUs} that execute and manage thousands of thread contexts at
the same time pose new challenges for employing helper threading,
which must be addressed carefully\comm{that prevents from direct usage
  of related prior work}. First, the numerous regular program threads
executing in parallel could require an equal or larger number of
helper threads that need to be managed at low cost. Second, the compute and
memory resources are dynamically partitioned between threads in GPUs,
and resource allocation for helper threads should be cognizant of
resource interference and overheads. Third, lock-step execution and
complex scheduling{\textemdash}which are characteristic of GPU
architectures{\textemdash}exacerbate the complexity of fine-grained
management of helper threads.

%In our work, we develop a new flexible framework (called \emph{Core-Assisted
%Bottleneck Acceleration} or \SADA) that dynamically identifies the
%underutilization of computational resources in modern GPUs, and uses these
%resources to accelerate the program executiom.
In this work, we describe a new, flexible framework for bottleneck
acceleration in GPUs via helper threading (called \emph{Core-Assisted
  Bottleneck Acceleration} or \SADA), which exploits the
aforementioned new opportunities while effectively handling the new
challenges.  \SADA performs acceleration by generating special
warps{\textemdash}\emph{\helperwarps}{\textemdash}that can execute
code to speed up application execution and system tasks. \comm{For example, \SADA can
  be used to perform compression/decompression of the data that is
  transferred between main memory and cores, thereby alleviating the
  off-chip memory bandwidth bottleneck.}To simplify the support of the
numerous assist threads with CABA, we manage their execution at the
granularity of a \emph{warp} and use a centralized mechanism to track
the progress of each \emph{assist warp} throughout its execution. To
reduce the overhead of providing and managing new contexts for each
generated thread, as well as to simplify scheduling and data
communication, an assist warp \emph{shares the same context} as the
regular warp it assists.  Hence, the regular warps are overprovisioned
with \emph{available registers} to enable each of them to host its own
assist warp.

%%% ONUR: make sure the compression I describe below is accurate and good.
\textbf{Use of CABA for compression.} We illustrate an important use case for
the CABA framework: alleviating the memory bandwidth bottleneck by enabling
\emph{flexible data compression} in the memory hierarchy.
%CABA enables the use of idle computational units and pipelines to perform
%efficient and flexible data compression.
The basic idea is to have assist warps that (1) compress cache blocks before
they are written to memory, and (2) decompress cache blocks before they are
placed into the cache.  

\green{\SADA-based compression/decompression provides several benefits
  over a purely hardware-based implementation of data compression for
  memory.} First, CABA primarily employs hardware that is already
available on-chip but is otherwise underutilized. In contrast,
hardware-only compression implementations require \emph{dedicated logic} for
specific algorithms.  Each new algorithm (or a modification of an
existing one) requires engineering effort and incurs hardware
cost. Second, different applications tend to have distinct data
patterns~\cite{bdi} that are more efficiently compressed with
different compression algorithms.\comm{Hence, application awareness in
  terms of choice of algorithm enables better exploitation of data
  compression.} CABA offers versatility in algorithm choice as we find
that many existing hardware-based compression algorithms (e.g.,
Base-Delta-Immediate (BDI) compression~\cite{bdi}, Frequent Pattern
Compression (FPC)~\cite{fpc}, and C-Pack~\cite{c-pack}) can be
implemented using different assist warps with the CABA
framework. Third, not all applications benefit from data
compression. Some applications are constrained by other bottlenecks
(e.g., oversubscription of computational resources), or may operate on
data that is not easily compressible.  As a result, the benefits of
compression may not outweigh the cost in terms of additional latency
and energy spent on compressing and decompressing data. In these
cases, compression can be easily disabled by CABA, and the CABA
framework can be used in other ways to alleviate the current
bottleneck.

%%% ONUR-final: Let's ignore the below paragraph. I like how it is
%%% written, but we will incorporate the wording at different places

%%% IGNORE
%\purple{This paper provides a detailed explanation and evaluation of
%  how CABA can be used to perform effective and flexible data
%  compression in GPU memory hierarchies.  We comprehensively evaluate
%  the performance and energy benefits of using CABA to enable
%  \emph{multiple} different compression algorithms and show that
%  significant performance and energy improvements are possible with
%  the CABA framework\comm{, when it is used for data compression}.}

%We discuss how CABA can enable the assist-warp based implementation of multiple
%compression algorithms without requiring dedicated hardware for them. 

\textbf{Other uses of CABA.} The generality of \green{CABA} enables
its use in alleviating other bottlenecks with different optimizations.
We discuss two examples: (1) \green{using assist warps to perform
  \emph{memoization}} to eliminate redundant computations that have
the same or similar inputs~\cite{dynreuse,Arnau-memo,danconnors}, by
storing the results of frequently-performed computations in the main
memory hierarchy (i.e., by converting the computational problem into a
storage problem) and, (2) \green{using the idle memory pipeline to
  perform opportunistic \emph{prefetching}} to better overlap
computation with memory access. Assist warps offer a hardware/software
interface to implement hybrid prefetching
algorithms~\cite{ebrahimi-hpca09} with varying degrees of complexity. We also
briefly discuss other uses of CABA for (1) redundant multithreading, (2)
speculative precomputation, (3) handling interrupts, and (4) profiling and
instrumentation.

%In this paper we make a case for CABA by providing a detailed study of how it
%can be used to perform efficient and flexible data compression in GPUs to
%reduce the memory bandwidth bottleneck.

\textbf{Contributions.} This work makes the following contributions:

%%% ONUR-final: I rephrased the contibutions to get the best of the
%%% removed paragraph and to make it more clear and well-flowing.

\begin{itemize}

\item It introduces the \emph{\SADAfull (\SADA) Framework}, which can
  mitigate different bottlenecks in modern GPUs by using
  underutilized system resources for \emph{assist warp} execution.

\item It provides a detailed description of how our framework can be
  used to enable effective and flexible data compression in GPU memory
  hierarchies.\comm{We show that \SADA can flexibly implement {\em
    multiple} different compression algorithms.}

%\comm{We exploit the benefits of \SADA's flexibility and
%    use it to implement multiple different compression algorithms
%    (BDI~\cite{bdi}, FPC~\cite{fpc}, and C-Pack~\cite{c-pack}).} We
%  observe that \SADA{\textemdash}when used to implement
%  compression{\textemdash}can reduce memory bandwidth consumption by
%  2.1X on average (up to 3.9X) across our workloads.

%%% ONUR: ``increase the effective on-chip bandwidth'' or ``reduce %% memory
%bandwidth consumption'' --> Can you use the latter if possible?

\item It comprehensively evaluates the use of \SADA for data
  compression to alleviate the memory bandwidth bottleneck. Our
  evaluations across a wide variety applications from
  Mars~\cite{mars}, CUDA~\cite{sdk}, Lonestar~\cite{lonestar}, and
  Rodinia~\cite{rodinia} \green{benchmark} suites show that
  \SADA-based compression on average (1) reduces memory bandwidth
  by 2.1X, (2) improves performance by 41.7\%, and (3)
  reduces overall system energy by 22.2\%.

\item It discusses at least six other use cases of CABA that can improve
application performance and system management, showing that CABA is a primary
general framework for taking advantage of underutilized resources in modern GPU
engines. 

\end{itemize}

\section{Background}
%\noindent{\bf GPU Architecture.}
\blue{A GPU consists of multiple simple cores, also called 
streaming multiprocessors (SMs) in NVIDIA terminology or compute units (CUs) in
AMD terminology. Our 
example architecture (shown in Figure~\ref{fig:architecture}) 
consists of 15 cores each with a SIMT width of 32, and 6 
memory controllers. Each core is associated with a private 
L1 data cache and read-only texture and constant caches 
along with a low latency, programmer-managed shared memory. 
The cores and memory controllers are connected via a crossbar 
and every memory controller is associated with a slice of the 
shared L2 cache. This architecture is similar to many modern GPU architectures,
including NVIDIA Fermi~\cite{fermi} and AMD Radeon~\cite{radeon}. 

\begin{figure}[h!]
\centering
\vspace{-0.2cm}
\includegraphics[width=0.35\textwidth]{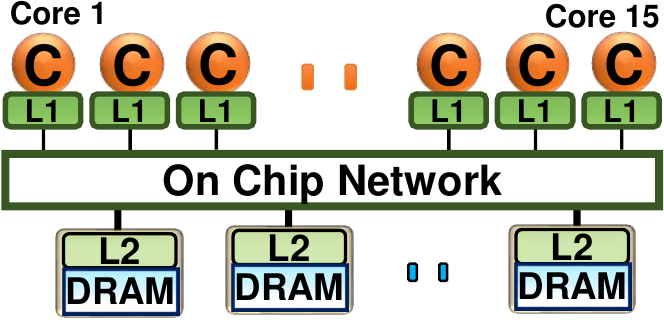}
\caption{Baseline GPU architecture. Figure reproduced from~\cite{caba}.}
\label{fig:architecture}
\vspace{-0.3cm}
\end{figure}

%\noindent{\bf Application Execution.}
A typical GPU application consists of many kernels. Each kernel is divided into groups
of threads, called thread-blocks (or cooperative thread arrays (CTAs)).
After a kernel is launched and its necessary data is copied to the GPU 
memory, the thread-block scheduler schedules available CTAs onto all the available cores~\cite{gpgpu-sim}.
Once the CTAs are launched onto the cores, the warps associated 
with the  CTAs are scheduled onto the cores' SIMT pipelines. Each core is capable of concurrently 
executing many warps, where a warp is typically defined 
as a group of threads that are executed in lockstep. In modern GPUs, a warp
can contain 32 threads~\cite{fermi}, for example. The maximum number of warps that 
can be launched on a core depends on the available 
core resources (e.g., available shared 
memory, register file size etc.).}
\comm{To efficiently support such fine-grain 
multi-threading, each core is provided with large register file to hold all 
the contexts of simultaneously active threads.}For example, in a modern GPU as many as 64 warps (i.e., 2048 threads) can be present on a single GPU
core. 

For more details on the internals of modern GPU architectures, we refer the
reader to ~\cite{wen-mei-hwu,patterson}.

\section{Motivation} \label{sec:motivation} 

We observe that different bottlenecks and imbalances
during program execution leave resources unutilized within the GPU
cores. \green{We motivate our proposal, CABA,  by examining these
inefficiencies.} CABA
leverages these inefficiencies as an opportunity to perform useful work.

\textbf{Unutilized Compute Resources.} A GPU core employs fine-grained
multithreading~\cite{burtonsmith,cdc6600} of {\em warps}, i.e., groups of threads executing the
same instruction, to hide long memory and ALU operation latencies.
\comm{When a few warps are stalled due to these long-latency
  operations, the remaining warps are swapped in for execution,
  potentially hiding the performance penalties of the stalled warps.
}If the number of available warps is insufficient to cover these long
latencies, the core stalls or becomes idle. \green{To understand} the
key sources of inefficiency in GPU cores, we conduct an experiment
where we show the breakdown of the applications' execution time spent
on either useful work (\emph{Active Cycles}) or stalling due to one of
four reasons: \green{\emph{Compute, Memory, Data Dependence
    Stalls} and \emph{Idle Cycles}}.  We also vary the amount of
available off-chip memory bandwidth: (i) half (1/2xBW), (ii) equal to
(1xBW), and (iii) double (2xBW) the peak memory bandwidth of our
baseline GPU architecture. Section~\ref{sec:methodology} details our baseline architecture and methodology.

\begin{figure*}[!h] \vspace{-0.5cm}
\centering \includegraphics[width=1\textwidth]{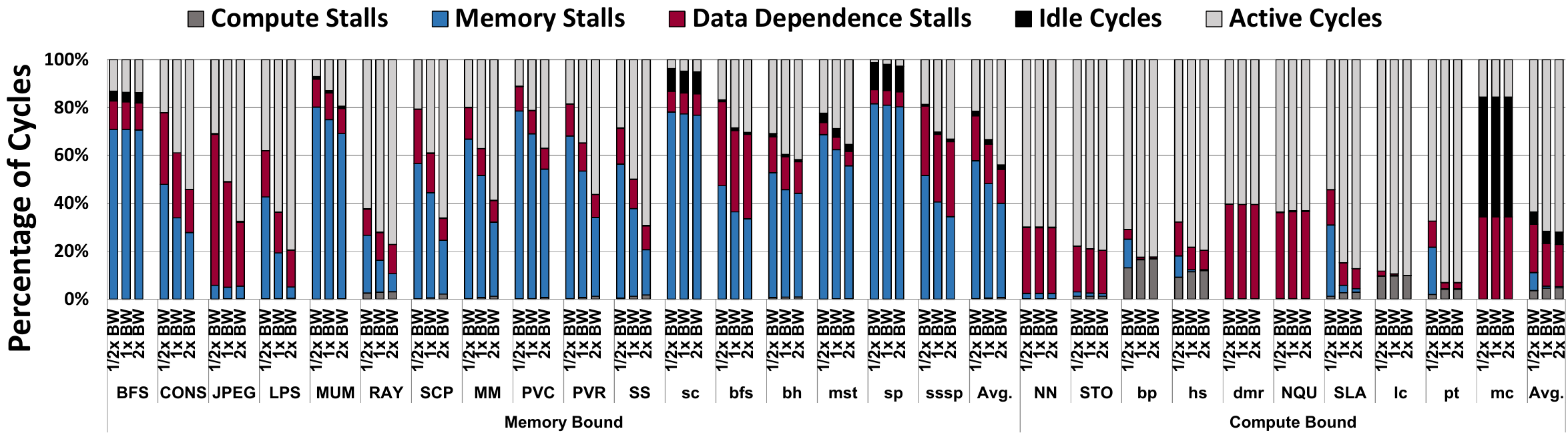}
\vspace{-0.4cm} \caption{\small{Breakdown of total issue cycles for 27 representative
CUDA applications. See Section~\ref{sec:methodology} for methodology. Figure
reproduced from~\cite{caba}.}}
\label{fig:pipelinestalls2} \vspace{-0.0cm} \end{figure*}

Figure~\ref{fig:pipelinestalls2} shows the percentage of total issue
cycles, divided into five components (as described above).  The
first two components{\textemdash}\emph{Memory and Compute Stalls}{\textemdash}are attributed
to the main memory and \green{ALU-pipeline structural stalls}. These stalls are because of backed-up
pipelines due to oversubscribed resources that \comm{situation arises
  due to the oversubscription to corresponding shared resources and
}prevent warps from being issued to the respective pipelines.  The
third component (\emph{Data Dependence Stalls}) is due to data
\green{dependence} stalls. These stalls prevent warps from
issuing new instruction(s) when the previous instruction(s) from the
same warp are stalled on long-latency operations (usually memory load
operations).  In some applications (e.g., {\tt dmr}),
special-function-unit (SFU) ALU operations that may take tens of
cycles to finish are also the source of data dependence stalls.  The
fourth component, \emph{Idle Cycles}, refers to idle cycles when either all
the available warps are issued to the pipelines and not ready
to execute their next instruction or the instruction buffers are flushed due to a mispredicted branch.  All these components are
sources of inefficiency that cause the cores to be underutilized. The
last component, \emph{Active Cycles}, indicates the fraction of cycles
during which at least one warp was successfully issued to the
pipelines.

We make two observations from Figure~\ref{fig:pipelinestalls2}.
First, \emph{Compute, Memory}, and \emph{Data Dependence Stalls} are
the major sources of underutilization in many GPU applications.  We distinguish applications
based on their primary bottleneck as either \emph{Memory} or
\emph{Compute Bound}. \green{We observe that a majority of the applications in our workload pool (17 out of
27 studied) are \emph{Memory Bound}, and bottlenecked by the off-chip memory bandwidth.}

\green{Second, for the \emph{Memory Bound} applications, we observe
  that the \emph{Memory} and \emph{Data Dependence} stalls constitute
  a significant fraction (61\%) of the total issue cycles on our
  baseline GPU architecture (1xBW). This fraction goes down to 51\%
  when the peak memory bandwidth is doubled (2xBW), and increases
  significantly when the peak bandwidth is halved (1/2xBW), indicating
  that limited off-chip memory bandwidth is a critical performance
  bottleneck for \emph{Memory Bound} applications. Some applications,
  e.g., \emph{BFS}, are limited by the interconnect bandwidth.
  \green{In contrast}, the \emph{Compute Bound} applications are
  primarily bottlenecked by stalls in the ALU pipelines\comm{that
    constitute 11\% of the underutilized cycles}.  An increase or
  decrease in the off-chip bandwidth has little effect on the
  performance of these applications.}

\begin{figure}[!h] \centering 
\vspace{-0.2cm} 
\centering \includegraphics[width=0.49\textwidth]{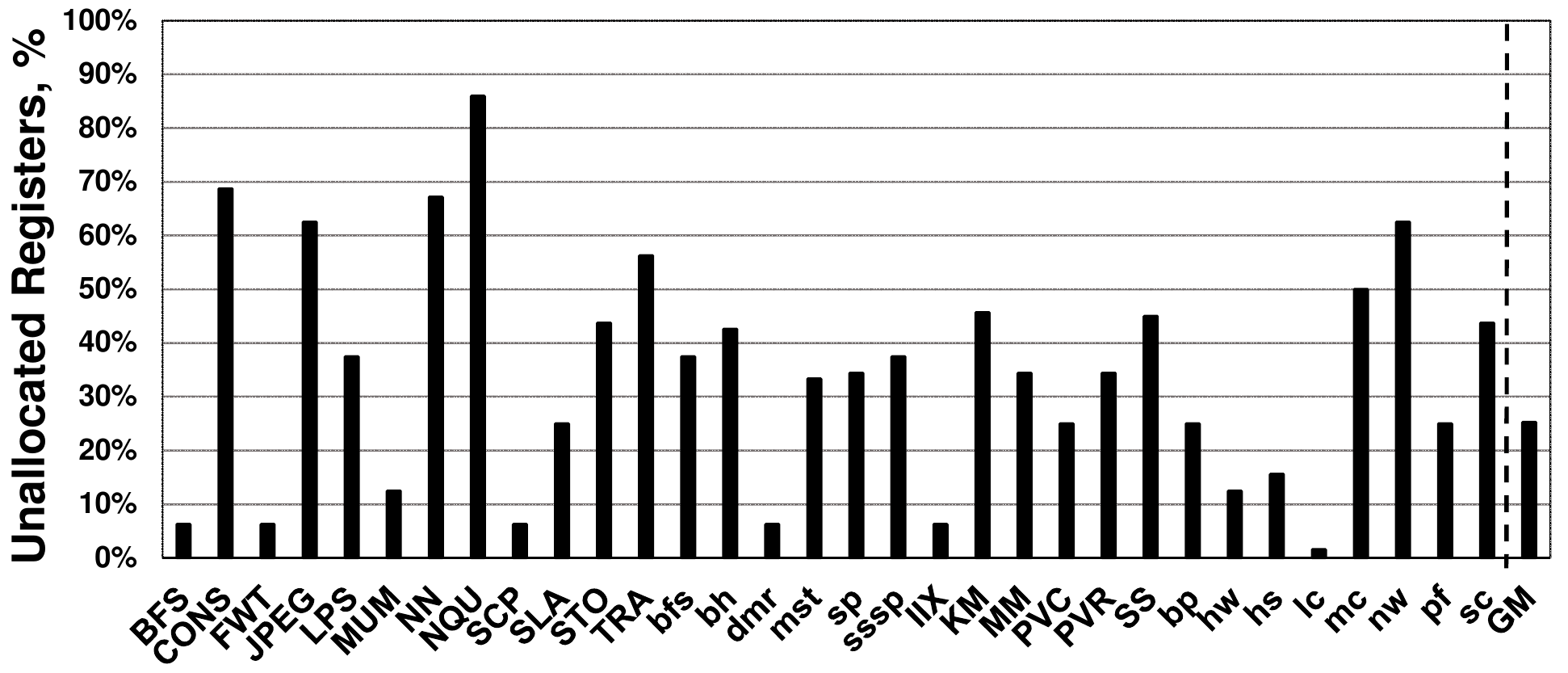}
\vspace{-0.4cm} \caption{Fraction of statically unallocated registers. Figure
reproduced from ~\cite{caba}.} \label{fig:reg_util}
\vspace{-0.2cm} \end{figure}

%%% ONUR-final2: threads/thread block --> Does this mean ``threads pe thread block'' or ``threads and thread blocks''. It is still unclear. Let's not use / when it is ambiguous what we mean.
%\vspace{-0.4cm} \vspace{-0.1cm}
\textbf{Unutilized On-chip Memory.} The \emph{occupancy} of any GPU Streaming
Multiprocessor (SM), i.e., the number of threads running concurrently,
is limited by a number of factors: (1) the available registers and
shared memory, (2) the hard limit on the number of threads and thread
blocks per core, (3) the number of thread blocks in the application
kernel. The limiting resource from the above, leaves the other resources
underutilized. This is because it is challenging, in practice, to achieve a
perfect balance in utilization of all of the above factors for different
workloads with varying characteristics. Very often, the factor determining the occupancy is the thread
or thread block limit imposed by the architecture. In this case, there
are many registers that are left unallocated to any thread
block. Also, the number of available registers may not be a multiple
of those required by each thread block. The remaining registers are
not enough to schedule an entire extra thread block, which leaves a
significant fraction of the register file and shared memory
unallocated and unutilized by the thread blocks.
Figure~\ref{fig:reg_util} shows the fraction of statically unallocated
registers in a 128KB register file (per SM) with a 1536 thread, 8
thread block occupancy limit, for different applications. We observe
that on average 24\% of the register file remains unallocated. This
phenomenon has previously been observed and analyzed in detail
in~\cite{unified-register, warped-register, spareregister,
  energy-register, compiler-register}. We observe a similar trend with
the usage of shared memory (not graphed).

\textbf{Our Goal.} We aim to exploit the underutilization of compute
resources, registers and on-chip shared memory as an opportunity to
enable different optimizations to accelerate various bottlenecks in
GPU program execution.\comm{For example, unutilized computational
  resources could be used to perform efficient data
  compression/decompression that effectively converts the data
  communication problem into a computational one.} To achieve this goal, we
would like to enable efficient helper threading for GPUs 
to dynamically generate threads in hardware \green{that use} the available
on-chip resources for various purposes. In the next section, we present the
detailed design of our \SADA framework that enables the generation
and management of these threads.\comm{The aim of the framework is to
  simplify the application of data compression as well as other
  optimization techniques (e.g., memoization) to accelerate the
  execution of GPU architectures.}
 
\comm{\begin{itemize} \item Detail what limits parallelism in GPUs:
\begin{itemize} \item CTAs/SM and threads/SM \item Program parallelism \item
Shared Memory/RF \end{itemize} \item Graph Showing unutilized shared memory and
register file \item Graph showing performance with increase in the threads/CTAs
per SM - to show oversubscription of bandwidth \item Discuss imbalance in
utilization of resources \item Point out that other papers have noticed this and
what they do about it.  \end{itemize}}

\section{The \SADA Framework} \label{sec:idea} In order to understand the major
design choices behind the \SADA framework, we first present our major design
goals and describe the key challenges in applying helper threading to GPUs. We
then show the detailed design, hardware changes, and operation of \SADA.
Finally, we briefly describe potential applications of our proposed framework.
Section \label{sec:compression} goes into a detailed design of one application
of the framework. 
 
\subsection{Goals and Challenges} 

%\begin{comment} Our goal in this work is to utilize available resources for
%useful computation.  In order to do that, we need to execute sub-routines that
%perform optimizations that accelerate bottlenecks in the application execution.
%The key difference between this form of assisted execution versus regular
%execution is that we want it to be \emph{low overhead} and in that sense they
%need to be treated differently from regular thread execution.  \comm{Supporting
%more regular program threads is non-trivial, requiring the different
%book-keeping structures to have increased capacity - e.g. warp schedulers,
%scoreboarding, branch-divergence support etc.  } The \emph{low overhead} goal
%leads us to a few requirements in designing a framework to enable helper
%threading.  We should be easily able to enable, trigger threads and kill them
%when required. We need them to be flexible enough to adapt to the run-time
%behavior of the regular program.  In addition to this, the helper thread needs
%to be able to communicate with the original thread. We need an interface to
%specify new subroutines and the framework should be generic enough to handle
%different optimizations.  \end{comment}

The purpose of \SADA is to leverage underutilized GPU resources for useful
computation.  To this end, we need to efficiently execute subroutines that
perform optimizations to accelerate \comm{application bottlenecks}bottlenecks in
application execution.  The key difference between \SADA's \emph{assisted
execution} and regular execution is that \SADA must be \emph{low overhead} and,
therefore, helper threads need to be treated differently from regular threads.
The \emph{low overhead} goal imposes several key requirements in designing a
framework to enable helper threading.  
%To achieve this efficiency, assisted execution should be done with \emph{low
%overhead} compared to normal (regular thread) execution.  This, in turn, leads
%to a few key requirements in designing a framework to enable helper threading.  
First, we should be able to easily manage helper threads{\textemdash}to enable,
trigger, and kill threads when required.  Second, helper threads need to be
flexible enough to adapt to the runtime behavior of the regular program.  Third,
a helper thread needs to be able to communicate with the original thread.
Finally, we need a flexible interface to specify new subroutines, with the
framework being generic enough to handle various optimizations.

%\comm{ In addition to addressing the above goals, applying helper threading to
%GPU architectures brings new challenges.  Execution on GPUs involves context
%switching between hundreds of threads. These threads are handled at different
%granularities in hardware and software. The programmer reasons about threads at
%the granularity of a thread block. Threads are scheduled onto GPU Streaming
%Multiprocessors at the granularity of a block but executed at any time at the
%granularity of a warp. \footnote{A warp is a group of threads that are executed
%in lock-step (32 in our configuration), while a thread block has upto 1024
%threads.} We need to decide the \emph{abstraction levels} for reasoning and
%managing helper threads from the point of view of the programmer, the hardware
%support as well as the compiler/runtime. Also, each executing thread could
%require execution of an associated helper thread routine.  Considering our goal
%of low overhead management, we need a mechanism to handle helper threads at
%this \emph{magnitude}.  }

With the above goals in mind, enabling helper threading in GPU architectures
introduces several new challenges.  First, execution on GPUs involves context
switching between hundreds of threads. These threads are handled at different
granularities in hardware and software. The programmer reasons about these
threads at the granularity of a thread block. \comm{Threads are scheduled onto
Streaming Multiprocessors (SMs) at the granularity of a block. }However, at any
point in time, the hardware executes only a small subset of the thread block,
i.e., a set of warps. Therefore, we need to define the \emph{abstraction
levels} for reasoning about and managing helper threads from the point of view
of the programmer, the hardware as well as the compiler/runtime. In addition,
each of the thousands of executing threads could simultaneously invoke an
associated helper thread subroutine.  To keep the management overhead low, we
need an efficient mechanism to handle helper threads at this magnitude.

Second, GPUs use fine-grained multithreading~\cite{cdc6600,burtonsmith} to
time multiplex the fixed number of compute units among the hundreds of threads.
Similarly, the on-chip memory resources (i.e., the register file and shared
memory) are statically partitioned between the different threads at compile
time. Helper threads require their own registers and compute cycles to
execute.  A straightforward approach would be to dedicate few registers and compute
units just for helper thread execution, but this option is both expensive and
wasteful. In fact, our primary motivation is to utilize \emph{existing idle
resources} for helper thread execution. In order to do this, we aim to enable
sharing of the existing resources between primary threads and helper threads at
low cost, while minimizing the interference to primary thread execution. In the
remainder of this section, we describe the design of our low-overhead \SADA
framework.

\subsection{Design of the \SADA Framework} \comm{\SADA enables helper threads, which we view as simple subroutines that
assist the primary program execution.} We choose to implement \SADA using a hardware/software co-design, as pure hardware or pure software approaches pose certain challenges that we describe below. There are two alternatives for a fully software-based
approach to helper threads.  The first alternative, treating each helper thread
as independent kernel code, has high overhead, since we are now treating the
helper threads as, essentially, regular threads. This would reduce the primary
thread occupancy in each SM (there is a hard limit on the number of
threads and blocks that an SM can support). It would also complicate the data communication
between the primary and helper threads, since no simple interface exists for
inter-kernel communication. The second alternative, embedding the
helper thread code within the primary thread kernel itself, offers little
flexibility in adapting to runtime requirements, since such helper threads cannot be triggered
or squashed independently of the primary thread. 

On the other hand, a pure
hardware solution would make register allocation for the assist warps and the
data communication between the helper threads and primary threads more difficult.
Registers are allocated to each thread block by the compiler and are then mapped to
the sections of the hardware register file at runtime. Mapping registers for
helper threads and enabling data communication between those registers and the primary thread
registers would be non-trivial.  Furthermore, a fully hardware approach would
make offering the programmer a flexible interface more challenging.

Hardware support enables simpler fine-grained management of helper threads,
aware of micro-architectural events and runtime program behavior.
Compiler/runtime support enables simpler context management for helper threads
and more flexible programmer interfaces. Thus, to get the best of both worlds, we propose a \emph{hardware/software cooperative approach}, where the hardware manages the scheduling and execution of helper thread subroutines, while the compiler
performs the allocation of shared resources (e.g., register file and shared
memory) for the helper threads and the programmer or the microarchitect provides the helper threads themselves. 

\subsubsection{Hardware-based management of threads.} To use the available
on-chip resources the same way that thread blocks do during program execution, we
dynamically insert sequences of instructions into the execution stream. We track
and manage these instructions at the granularity of a warp, and refer to them as
\emph{\textbf{Assist Warps}}. An \helperwarp is a set of
instructions issued into the core pipelines. Each instruction is executed in
lock-step across all the SIMT lanes, just like any regular instruction, with an
active mask to disable lanes as necessary. The \helperwarp does \emph{not} own a
separate context (e.g., registers, local memory), and instead shares both a
context and a warp ID with the regular warp that invoked it. In other words, each
assist warp is coupled with a \emph{parent warp}. In this sense, it is different
from a regular warp and does not reduce the number of threads that
can be scheduled on a single SM. Data sharing between the two warps becomes simpler, since the assist warps share the register file with the parent warp.
Ideally, an assist warp consumes resources and issue cycles that would otherwise
be idle. We describe the structures required to support hardware-based
management of assist warps in Section~\ref{sec:Components}. 

\subsubsection{Register file/shared memory allocation.} Each helper thread
subroutine requires a different number of registers depending on the
actions it performs. These registers have a short lifetime, with no
values being preserved between different invocations of an assist warp. To limit the register requirements for assist warps, we impose the restriction
that only one instance of each helper thread routine can be active for each
thread. All instances of the same helper thread for each parent thread use
the same registers, and the registers are allocated to the helper threads statically by the compiler. One of
the factors that determines the runtime SM occupancy is the number of registers required by a thread block (i.e, per-block register requirement). For each helper thread subroutine that is enabled, we add its register requirement
to the per-block register requirement, to ensure the availability of registers
for both the parent threads as well as every assist warp. The
registers that remain unallocated after allocation among the parent thread
blocks should suffice to support the assist warps. If not, register-heavy assist
warps may limit the parent thread block occupancy in SMs or increase the number
of register spills in the parent warps. Shared memory resources are partitioned in a similar manner and allocated to each assist warp as and if
needed.  

\subsubsection{Programmer/developer interface.} The assist warp subroutine can
be written in two ways. First, it can be supplied and annotated by the
programmer/developer using CUDA extensions with PTX instructions and then
compiled with regular program code. Second, the assist warp subroutines can be written by the microarchitect in the internal GPU instruction format. \comm{This approach enables the
microarchitect to take control at finer granularity, and }These helper thread
subroutines can then be enabled or disabled by the application programmer. This
approach is similar to that proposed in prior work (e.g.,~\cite{ssmt}). It
offers the advantage of potentially being highly optimized for energy and
performance while having flexibility in implementing optimizations
that are not trivial to map using existing GPU PTX instructions.  The
instructions for the helper thread subroutine are stored in an on-chip buffer
(described in Section~\ref{sec:Components}). 

Along with the helper thread subroutines, the programmer also provides: (1)
the \emph{priority} of the assist warps to enable the warp scheduler to
make informed decisions,  (2) the trigger conditions for each assist warp, and (3)
the live-in and live-out variables for data communication with the parent warps.

Assist warps can be scheduled with different priority levels in relation to
parent warps by the warp scheduler. Some assist warps may perform a function
that is required for correct execution of the program and are \emph{blocking}.
At this end of the spectrum, the \emph{high priority} assist warps are treated
by the scheduler as always taking higher precedence over the parent warp
execution. Assist warps
should be given a high priority only when they are \comm{on the critical path for
program execution and are }required for correctness. \emph{Low priority} assist
warps, on the other hand, are scheduled for execution only when computational
resources are available, i.e., during idle cycles. There is no guarantee that
these assist warps will execute or complete. 

The programmer also
provides the conditions or events that need to be satisfied for the deployment of
the assist warp. This includes a specific point within the original program
and/or a set of other microarchitectural events that could serve as a \emph{trigger}
for starting the execution of an assist warp.\comm{With assist warps that are pre-installed by the
microarchitect, the trigger events and priority are hard-coded for each assist
warp routine, and the programmer is provided with a knob to enable/disable a
specific assist warp routine.}

\subsection{Main Hardware Additions} \label{sec:Components}
\begin{figure}[h] \centering \vspace{-0.1cm}
\includegraphics[width=0.49\textwidth]{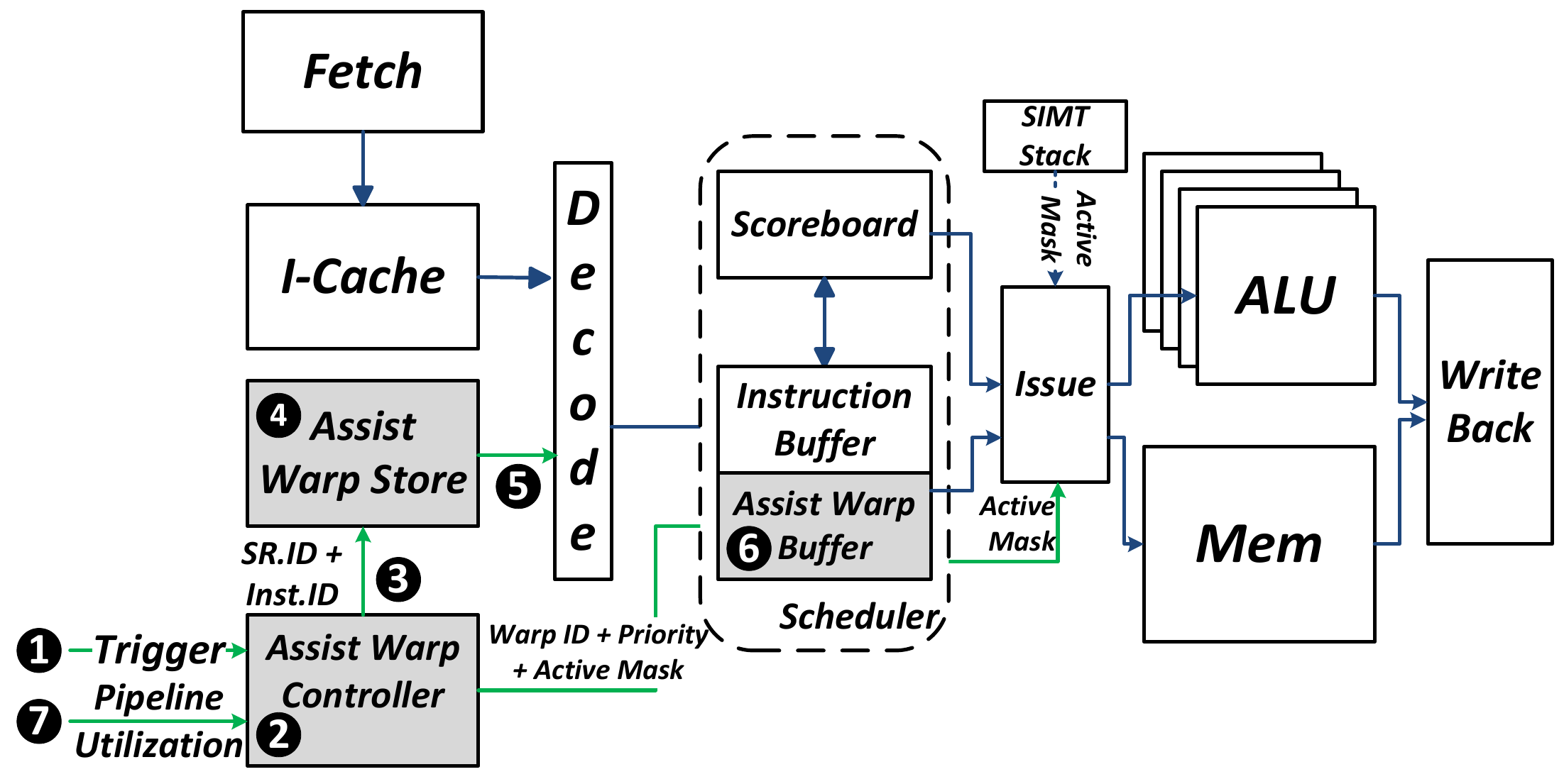} \vspace{-0.5cm}
\caption{\SADA framework flow within a typical GPU pipeline~\cite{manual}. The
shaded blocks are the components introduced for the framework. Figure reproduced
from~\cite{caba}.}
\label{fig:framework} \vspace{-0.2cm} \end{figure}

Figure~\ref{fig:framework} shows a high-level block diagram of the GPU
pipeline~\cite{manual}.\comm{Instructions are fetched from the Instruction
cache, decoded and then buffered in the Instruction Buffer before a warp is
selected by the scheduler for issue into the ALU/memory pipelines. Accesses to
global memory also involve address generation and coalescing before requests are
send to the memory hierarchy.} To support assist warp execution, we add three new
components: (1) an Assist Warp Store to hold the assist warp
code, (2) an Assist Warp Controller to perform the deployment, tracking,
and management of assist warps, and (3) an Assist Warp Buffer to stage instructions from triggered assist
warps for execution.

\textbf{Assist Warp Store (AWS).} Different assist warp subroutines are possible
based on the purpose of the optimization. These code sequences for different
types of \helperwarps need to be stored on-chip. An on-chip storage structure
called the Assist Warp Store (\ding{205}) is preloaded with these instructions
before application execution. It is indexed using the subroutine index (SR.ID)
along with the instruction ID (Inst.ID).\comm{Indexing is based on the sequence
number. Special instructions are required to load the \helperwarp sequencer
before the program execution begins.}\comm{Note that an alternative
approach, is to simply contain the AWS as a part of the instruction cache. We, however, assume a separate structure to carry assist warp code.}

\textbf{Assist Warp Controller (AWC).} The AWC (\ding{203}) is responsible for the triggering, tracking, and management
of \helperwarp execution. It stores a mapping between
trigger events and a subroutine index in the AWS, as specified by the programmer. The AWC monitors for such events, and when they take place, triggers the fetch, decode and execution of instructions from the AWS for the respective assist warp. 

Deploying all the instructions within an assist warp, back-to-back, at the trigger point may require increased fetch/decode bandwidth and buffer space after decoding~\cite{ssmt2}. To avoid this, at each
cycle, only a few instructions from an assist warp, at most equal to the
available decode/issue bandwidth, are decoded and staged for execution. Within the AWC, we simply track the next instruction that needs to be executed for each assist warp and this is stored in the Assist Warp Table (AWT), as depicted in Figure~\ref{fig:AWT}. The AWT also tracks additional metadata required for assist warp management, which is described in more detail in Section~\ref{sec:mechanism}.  

\textbf{Assist Warp Buffer (AWB).} Fetched and decoded instructions (\ding{203}) belonging to the assist warps that have been triggered need to
be buffered until the assist warp can be selected for issue by the scheduler. These instructions are then staged in the Assist
Warp Buffer (\ding{207}) along with their warp IDs. The AWB is contained within the \emph{instruction buffer (IB)}, which holds decoded instructions for the parent warps. The AWB makes use of the existing IB structures. The IB is typically partitioned among different warps executing in the SM. Since each assist warp is associated with a parent warp, the assist warp instructions are directly inserted into the \emph{same partition} within the IB as that of the parent warp. This simplifies warp scheduling, as the assist warp instructions can now be issued as if they were parent warp instructions with the same warp ID. In addition, using the existing partitions avoids the cost of separate dedicated instruction buffering for assist warps. We do, however,
provision a small additional partition with two entries within the IB, to hold
non-blocking \emph{low priority} assist warps that are scheduled only during idle
cycles. This additional partition allows the scheduler to distinguish \emph{low priority} assist warp instructions from the parent warp and \emph{high priority} assist warp instructions, which are given precedence during scheduling, allowing them to make progress. 

\subsection{The Mechanism}
\label{sec:mechanism}

\begin{figure}[t] \centering \vspace{-0.0cm}
\includegraphics[width=0.49\textwidth]{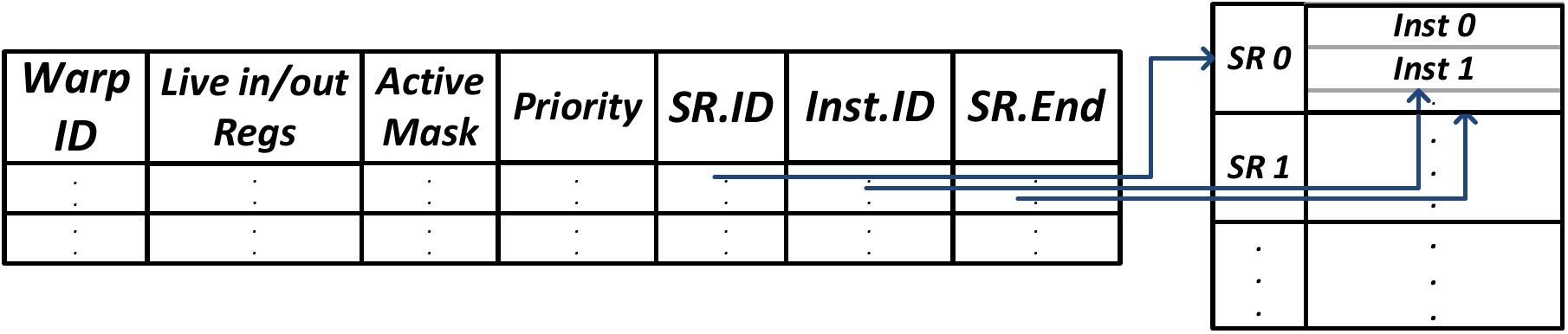} \vspace{-0.4cm}
\caption{Fetch Logic: Assist Warp Table (contained in the AWC) and the Assist
Warp Store (AWS). Figure reproduced from~\cite{caba}.} \label{fig:AWT} \vspace{-0.1cm} \end{figure}

\indent\textbf{Trigger and Deployment.} An \helperwarp is triggered (\ding{202}) by the AWC (\ding{203}) based on a specific
set of architectural events and/or a triggering instruction (e.g., a load
instruction).  When an \helperwarp is triggered, its specific
instance is placed into the Assist Warp Table (AWT) within the AWC
(Figure~\ref{fig:AWT}). Every cycle, the AWC selects an assist warp to deploy in
a round-robin fashion. The AWS is indexed (\ding{204}) based on the subroutine
ID (SR.ID){\textemdash}which selects the instruction sequence to be executed by the assist
warp, and the instruction ID (Inst.ID){\textemdash}which is a pointer to the next instruction
to be executed within the subroutine (Figure~\ref{fig:AWT}). The selected
instruction is entered (\ding{206}) into the AWB (\ding{207}) \comm{The AWC also generates
the priority of each \helperwarp to help the scheduler prioritize between the
\helperwarps and primary warps. }and, at this point, the instruction enters the active
pool with other active warps for scheduling. The Inst.ID for the assist warp is updated in the AWT to point to the next
instruction in the subroutine. When the end of the subroutine is reached, the
entry within the AWT is freed. 

\textbf{Execution.} Assist warp instructions, when selected for issue by the scheduler, are executed in much the same way as any other instructions. The scoreboard tracks the dependencies between instructions
within an assist warp in the same way as any warp, and instructions from different assist warps are interleaved in execution in order to hide latencies. \purple{We also provide an active mask (stored as a part of the AWT), which 
allows for statically disabling/enabling different lanes within a warp. This is
useful to provide flexibility in lock-step instruction execution when we do not need all threads within a warp to execute a specific assist warp subroutine.} 

\textbf{Dynamic Feedback and Throttling.} \Helperwarps, if not properly
controlled, may stall application execution. This can happen due to several
reasons.  First, \helperwarps take up issue cycles, and only a limited number of
instructions may be issued per clock cycle.  Second, \helperwarps require
structural resources: the ALU units and resources in the load-store pipelines
(if the assist warps consist of computational and memory instructions,
respectively). We may, hence, need to throttle assist warps
to ensure that their performance benefits outweigh the overhead. This requires mechanisms to appropriately balance and manage the
aggressiveness of assist warps at runtime. 

The overheads associated with \helperwarps can be controlled in different ways.
First, the programmer can statically specify the priority of the \helperwarp.
Depending on the criticality of the \helperwarps in making forward progress,
the assist warps can be issued either in idle cycles or with varying levels of priority in
relation to the parent warps.  For example, warps performing \emph{decompression} are given a
high priority whereas warps performing \emph{compression} are given a low
priority. Low priority assist warps are inserted into the dedicated
partition in the IB, and are scheduled only during idle cycles. This priority is
statically defined by the programmer. Second, the AWC can control the number of times
the \helperwarps are deployed into the AWB.\comm{In our framework applied to data compression, we
employ these options depending on the overhead associated with the
\helperwarps.}
The AWC monitors the utilization of the functional
units (\ding{208}) and idleness of the cores to decide when to throttle assist warp deployment. \comm{A simple
timeout mechanism is used by the AWC to periodically flush lower priority assist warps from
the Assist Warp Table if the AWC is unable to find idle cycles to issue them.} 

\textbf{Communication and Control.} An assist warp may need to communicate data
and status
with its parent warp. For example, memory addresses from the parent warp need to
be communicated to assist warps performing decompression or prefetching. The IDs
of the registers containing the live-in data for each assist warp are saved in the AWT
when an assist warp is triggered. Similarly, if an assist warp needs to report
results to its parent warp (e.g., in the case of memoization), the register
IDs are also stored in the AWT. When the assist warps execute,
\emph{MOVE} instructions are first executed to copy the live-in data from the
parent warp registers to the assist warp registers. Live-out
data is communicated to the parent warp in a similar fashion, at the end of
assist warp execution.

Assist warps may need to be \emph{killed} when they are not required (e.g., if the data
does not require decompression) or when they are no longer beneficial. In this case, the entries in the AWT and AWB are
simply flushed for the assist warp. 

%We implement the \SADA Framework in GPGPU-Sim Version 3.1~\cite{GPGPUSim} and
\subsection{Applications of the \SADA Framework} We envision multiple
applications for the \SADA framework, e.g., data
compression~\cite{fvc,fpc,c-pack,bdi}, memoization~\cite{dynreuse, Arnau-memo,
danconnors}, data prefetching~\cite{BPKI, stream1,stream2,stride1,stride2}. In Section~\ref{sec:compression}, we provide a detailed case study of enabling data compression with
the framework, discussing various tradeoffs. We
believe \SADA can be useful for many other optimizations, and we discuss some of them
briefly in Section~\ref{sec:Applications}. 
%In this work, we first discuss some optimizations that can be implemented using
%assist warps. We then apply the \SADA framework to enable and evaluate
%\emph{\textbf{Data Compression}} as a means to mitigate the memory bandwidth
%bottleneck.\footnote{We implement and evaluate the \SADA Framework in GPGPU-Sim
%v3.2.1~\cite{GPGPUSim}.} 

\section{A Case for CABA: Data Compression}
\label{sec:compression}
%\comm{In this paper, we evaluate \emph{Data Compression} using the
%\emph{Base-Delta-Immediate compression} algorithm.% in bandwidth constrained
%workloads.  We now briefly describe the compression algorithm and its
%implementation, and then delve into mechanism details and micro-architectural
%changes required for core-assisted data compression and decompression.
%Finally, we discuss optimizations to reduce the performance/power overhead of
%\SADA and improve its versatility.} 
Data compression is a technique that exploits the redundancy in the
applications' data to reduce capacity and bandwidth requirements for many modern
systems by saving and transmitting data in a more compact form.  Hardware-based
data compression has been explored in the context of on-chip
caches~\cite{fvc,fpc,c-pack,bdi,dcc,sc2,zca,zvc,camp},
interconnect~\cite{noc-comp}, and main
memory~\cite{MXT,LinkCompression,MMCompression,lcp-micro,memzip,toggle-aware-cal,toggle-aware-hpca} as a
means to save storage capacity as well as memory bandwidth.  In modern GPUs,
memory bandwidth is a key limiter to system performance in many workloads
(Section~\ref{sec:motivation}). As such, data compression is a promising
technique to help alleviate this bottleneck.  Compressing data enables less data
to be transferred from/to DRAM and the interconnect. 

In
bandwidth-constrained workloads, idle compute pipelines offer an opportunity to
employ \SADA to enable data compression in GPUs. We can use assist warps to (1)
decompress data, before loading it into the caches and registers, and (2)
compress data, before writing it back to memory. Since assist warps execute
instructions, \SADA offers some flexibility in the compression algorithms that
can be employed. Compression algorithms that can be mapped to the general GPU
execution model can be flexibly implemented with the \SADA framework. \comm{This
section describes how we can use the CABA framework to implement data
compression in GPUs.}  

\subsection{Mapping Compression Algorithms into Assist Warps} 

In order to employ \SADA to enable data compression, we need to map compression
algorithms into instructions that can be executed within the GPU cores. For a
compression algorithm to be amenable for implementation with \SADA, it ideally
needs to be (1) reasonably parallelizable and (2) simple (for low latency).
Decompressing data involves reading the encoding associated with each cache line
that defines how to decompress it, and then triggering the corresponding
decompression subroutine in CABA.\comm{The instruction storage in the AWS
constrains the number of possible encodings. } Compressing data, on the other
hand, involves testing different encodings and saving data in the compressed
format. 

We perform compression at the granularity of a cache line. The data needs to be
decompressed before it is used by any program thread.  In order to utilize the
full SIMD width of the GPU pipeline, we would like to decompress/compress all
the words in the cache line in parallel. With \SADA, helper thread routines are
managed at the warp granularity, enabling fine-grained triggering of assist
warps to perform compression/decompression when required. However, the SIMT
execution model in a GPU imposes some challenges: (1) threads within a warp
operate in lock-step, and (2) threads operate as independent entities, i.e.,
they do not easily communicate with each other. 

In this section, we discuss the architectural changes and algorithm adaptations
required to address these challenges and provide a detailed implementation and
evaluation of \emph{Data Compression} within the \SADA framework using the
\emph{Base-Delta-Immediate compression} algorithm~\cite{bdi}.
%We briefly describe the compression and decompression algorithms, their
%mappings to the GPU, and then delve into mechanism details and
%micro-architectural changes required for core-assisted data compression and
%decompression. 
Section~\ref{sec:Other_Algos} discusses implementing other compression
algorithms.

\subsubsection{Algorithm Overview.} \label{sec:algorithm} 
\begin{figure}[!h] \vspace{-0.1cm}
\includegraphics[scale=0.35]{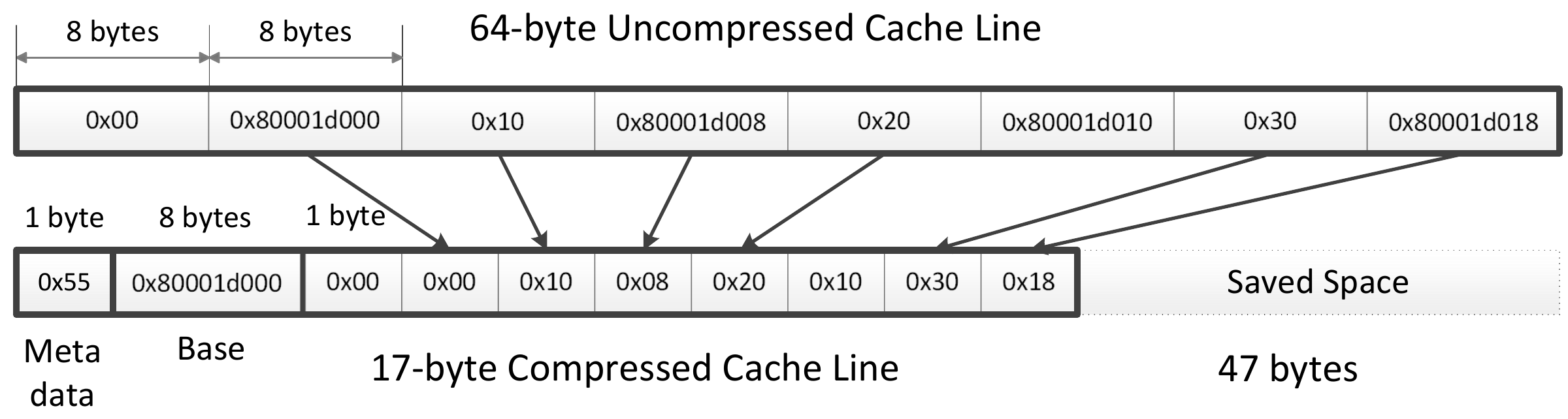} \vspace{-0.1cm}
\caption{Cache line from \emph{PVC} compressed with BDI. Figure reproduced
from~\cite{caba}.} \vspace{-0.1cm}
\label{fig:bdc-example2} \end{figure}

Base-Delta-Immediate
compression (BDI) is a simple compression algorithm that was originally proposed
in the context of caches~\cite{bdi}.  It is based on the observation that many
cache lines contain data with low dynamic range. BDI exploits this observation
to represent a cache line with low dynamic range using a common \emph{base} (or
multiple bases) and an array of \emph{deltas} (where a delta is the difference
of each value within
the cache line and the common base). Since the \emph{deltas} require fewer bytes
than the values themselves, the combined size after compression can be much
smaller. Figure~\ref{fig:bdc-example2} shows the compression of an example
64-byte cache line from the \emph{PageViewCount (PVC)} application using BDI.
As Figure~\ref{fig:bdc-example2} indicates, in this case, the cache line can be
represented using two bases (an 8-byte base value, $0x8001D000$, and an implicit
zero value base) and an array of eight 1-byte differences from these bases. As a
result, the entire cache line data can be represented using 17 bytes instead of
64 bytes (1-byte metadata, 8-byte base, and eight 1-byte deltas), saving 47
bytes of the originally used space.

Our example implementation of the BDI compression algorithm~\cite{bdi} views a
cache line as a set of fixed-size values i.e., 8 8-byte, 16 4-byte, or 32 2-byte
values for a 64-byte cache line.  For the size of the deltas, it considers three
options: 1, 2 and 4 bytes.  \comm{Note that all potential compressed sizes are
known statically after compression.} The key characteristic of BDI, which makes
it a desirable compression algorithm to use with the \SADA framework, is its
fast parallel decompression that can be efficiently mapped into instructions
that can be executed on GPU hardware. Decompression is simply a masked vector
addition of the deltas to the appropriate bases~\cite{bdi}.

\subsubsection{Mapping BDI to CABA.} In order to implement BDI with the \SADA
framework, we need to map the BDI compression/decompression algorithms into GPU
instruction subroutines (stored in the AWS and deployed as \helperwarps).

\textbf{Decompression.} To decompress the data compressed with BDI, we need a
simple addition of deltas to the appropriate bases. The CABA decompression
subroutine first loads the words within the compressed cache line into assist
warp registers, and then performs the base-delta additions in parallel, employing the
wide ALU pipeline.\footnote{Multiple instructions are required if the number of
deltas exceeds the width of the ALU pipeline. We use a 32-wide pipeline.} The
subroutine then writes back the uncompressed cache line to the cache. It skips
the addition for the lanes with an implicit base of zero by updating the active
lane mask\comm{manipulating the per-lane predicate registers~\cite{pguide}}
based on the cache line encoding.  We store a separate subroutine for each
possible BDI encoding that loads the appropriate bytes in the cache line as the
base and the deltas. \blue{The high-level algorithm for decompression is
presented in Algorithm 1.}

%\begin{document}
\vspace{-0.1cm} \begin{algorithm} \caption{BDI: Decompression} \small{
\begin{algorithmic}[1] \State load \emph{base, deltas} \State
\emph{uncompressed\_data} $=$ \emph{base} $+$ \emph{deltas} \State store
\emph{uncompressed\_data} \end{algorithmic} \vspace{-0.1cm} } \end{algorithm} 
%\end{document}

\textbf{Compression.} To compress data, the \SADA compression subroutine tests
several possible encodings (each representing a different size of base and
deltas) in order to achieve a high compression ratio. The first few bytes (2--8
depending on the encoding tested) of the cache line are always used as the base.
\purple{Each possible encoding is tested to check whether the cache line can be
successfully encoded with it. In order to perform compression at a warp
granularity, we need to check whether all of the words at every SIMD lane were
successfully compressed. In other words, if any one word cannot be compressed,
that encoding cannot be used across the warp. We can perform this check by
adding a global predicate register, which stores the logical AND of the per-lane
predicate registers.\comm{Each option is tested to check whether the line is
compressible using the specific encoding. The compressibility of all lanes for a
specific encoding, in parallel, is detected using a global predicate register.}
We observe that applications with homogeneous data structures can typically use the same
encoding for most of their cache lines~\cite{bdi}. We use this observation to reduce the
number of encodings we test to just one in many cases.} \comm{In BDI, the
number of instructions required to perform the compression depends on the
compressibility of the data.  If more encodings are tested, a better compression
ratio can be obtained at the cost of higher latency. The encoding complexity can
be decided based on the requirement at run time. If we just use one encoding
this would translate to approximately three operations. After compression, the
first few bytes (2--8 depending on the encoding used) of the cache line are
always used as the base.} All necessary operations are done in parallel using
the full width of the GPU SIMD pipeline.
\blue{The high-level algorithm for compression is presented in Algorithm 2.}

\begin{algorithm} \caption{BDI: Compression} \small{
\begin{algorithmic}[1] \For{\emph{each base\_size}} \State load \emph{base,
values} \For{\emph{each delta\_size}} \State \emph{deltas} $=$ \emph{abs(values -
base)} \If{\emph{size(deltas) $<=$ delta\_size}} \State store \emph{base, deltas}
\State \textbf{exit} \EndIf \EndFor \EndFor \end{algorithmic} } \end{algorithm}

\subsubsection{Implementing Other Algorithms.} \label{sec:Other_Algos} The BDI
compression algorithm is naturally amenable towards implementation using assist
warps because of its data-parallel nature and simplicity.  The CABA framework
can also be used to realize other algorithms. The challenge in implementing
algorithms like FPC~\cite{fpc-tr} and C-Pack~\cite{c-pack}\comm{Our
technical report~\cite{caba-tr} and the original works~\cite{fpc-tr,c-pack}
provide more details on the specifics of these algorithms.}, which have
variable-length compressed words, is primarily in the placement of compressed
words within the compressed cache lines.  In BDI, the compressed words are in
\emph{fixed} locations within the cache line and, for each encoding, all the
compressed words are of the same size and can, therefore, be processed in
parallel. In contrast, C-Pack may employ multiple dictionary values as opposed
to just one base in BDI. In order to realize algorithms with \emph{variable
length words} and \emph{dictionary values} with assist warps, we leverage the
coalescing/address generation logic~\cite{coal1,coal2} already available in the
GPU cores. We make two minor modifications to these
algorithms~\cite{fpc-tr,c-pack} to adapt them for use with \SADA. First, similar
to prior works~\cite{c-pack,fpc-tr,MMCompression}, we observe that few encodings
are sufficient to capture almost all the data redundancy. In addition, the
impact of any loss in compressibility due to fewer encodings is minimal as the
benefits of bandwidth compression are only at multiples of a single DRAM burst
(e.g., 32B for GDDR5~\cite{GDDR5}). We exploit this to reduce the number of
supported encodings.  \comm{reduce the number of encodings supported, as we
observe that just a few encodings are sufficient to capture almost all of the
redundancy in data.\footnote{Prior works~\cite{c-pack,fpc-tr,MMCompression} made
a similar observation.}} Second, we place all the metadata containing the
compression encoding at the \emph{head} of the cache line to be able to determine how
to decompress the entire line \emph{upfront}. In the case of C-Pack, we place the
dictionary entries after the metadata.  

We note that it can be challenging to implement complex algorithms efficiently
with the simple computational logic available in GPU cores.  Fortunately, there
are already Special Function Units (SFUs)~\cite{sfu,sfu2} present in the GPU
SMs, used to perform efficient computations of elementary mathematical
functions.  SFUs could potentially be extended to implement primitives that
enable the fast iterative comparisons performed frequently in some compression
algorithms. This would enable more efficient execution of the described
algorithms, as well as implementation of more complex compression algorithms,
using \SADA. We leave the exploration of an SFU-based approach to future work. 

\blue{We now present a detailed overview of mapping the FPC and C-PACK algorithms into
assist warps.}

\subsubsection{Implementing the FPC (Frequent Pattern Compression)
Algorithm.} For FPC, the cache line is treated as set of
fixed-size words and each word within the cache line is compressed into a simple
{prefix or encoding and a compressed word if it matches a set of frequent
patterns, e.g. narrow values, zeros or repeated bytes. The word is left
uncompressed if it does not fit any pattern. We refer the reader to the original
work~\cite{fpc-tr} for a more detailed description of the original algorithm.

The challenge in mapping assist warps to the FPC
decompression algorithm is in the serial sequence in which each word within a
cache line is decompressed. This is because in the original proposed version,
each compressed word can have a different size. To determine the location of a
specific compressed word, it is necessary to have decompressed the previous
word. We make some modifications to the algorithm in order to parallelize the decompression across
different lanes in the GPU cores. First, we move the word prefixes (metadata)
for each word to the front of the cache line, so we know \emph{upfront} how to
decompress the rest of the cache line. Unlike with BDI, each word within the
cache line has a different encoding and hence a different compressed word length
and encoding pattern. This is problematic as statically storing the sequence of
decompression instructions for every combination of patterns for all the words
in a cache line would require very large instruction storage. In order to
mitigate this, we break each cache line into a number of segments. Each segment
is compressed independently and all the words within each segment are compressed
using the \emph{same encoding} whereas different segments may have different
encodings. This creates a trade-off between
simplicity/parallelizability versus compressibility. Consistent with
previous works~\cite{fpc-tr}, we find that this doesn't significantly impact compressibility. 

\textbf{Decompression.} The high-level algorithm we use for decompression is presented
in Algorithm 3. Each segment within the compressed cache line is loaded in series. Each
of the segments is decompressed in parallel{\textemdash}this is possible because
all the compressed words within the segment have the same encoding. The
decompressed segment is then stored before moving onto the next segment. The
location of the next compressed segment is computed based on the size of the
previous segment.

 \begin{algorithm} \caption{FPC: Decompression} \small{ \begin{algorithmic}[1] \For{\emph{each segment}}
\State load \emph{compressed words} \State \emph{pattern specific decompression (sign
extension/zero value)} \State store \emph{decompressed words} \State
\emph{segment-base-address} $=$ \emph{segment-base-address} $+$ \emph{segment-size}
\EndFor \end{algorithmic}}  \end{algorithm} 

\textbf{Compression.} Similar to the BDI implementation, we loop through and test
different encodings for each segment. We also compute the address offset for each
segment at each iteration to store the compressed words in the appropriate
location in the compressed cache line. Algorithm 4 presents the high-level FPC
compression algorithm we use. 
 
\begin{algorithm} \caption{FPC: Compression} \small{ \begin{algorithmic}[1]
\State load \emph{words}
\For{\emph{each segment}}  \For{\emph{each encoding}}  \State \emph{test
encoding} \If{\emph{compressible}} \State \emph{segment-base-address} $=$
\emph{segment-base-address} $+$ \emph{segment-size} \State store \emph{compressed
words} \State \textbf{break} \EndIf \EndFor \EndFor \end{algorithmic} }
\end{algorithm}

\textbf {Implementing the C-Pack Algorithm.} C-Pack~\cite{c-pack} is a
dictionary based compression
algorithm where frequent "dictionary" values are saved at the beginning of the
cache line. The rest of the cache line contains encodings for each word which may
indicate zero values, narrow values, full or partial matches into the
dictionary or any word that is uncompressible. 

In our implementation, we reduce
the number of possible encodings to partial matches (only last byte mismatch),
full word match, zero value and zero extend (only last byte) and we limit the
number of dictionary values to 4. This enables fixed compressed word size within
the cache line. A fixed compressed word size enables compression and decompression
of different words within the cache line in parallel. If the number of required
dictionary values or uncompressed words exceeds 4, the line is left
decompressed. This is, as in BDI and FPC, a trade-off between simplicity and
compressibility. In our experiments, we find that it does not significantly
impact the compression ratio{\textemdash}primarily due the 32B minimum data size
and granularity of compression. 

\textbf{Decompression.} As described, to enable parallel decompression, we place the
encodings and dictionary values at the head of the line. We also limit the
number of encodings to enable quick
decompression. We implement C-Pack decompression as a series of instructions
(one per encoding used) to load all the registers with the appropriate
dictionary values. We define the active lane mask based on the encoding (similar
to the mechanism used in BDI) for each load instruction to ensure the correct
word is loaded into each lane's register. Algorithm 5 provides the high-level
algorithm for C-Pack decompression.  

\begin{algorithm} \caption{C-PACK: Decompression} \small{ \begin{algorithmic}[1] \State add \emph{base-address}
$+$ \emph{index-into-dictionary} \State load \emph{compressed words} \For{each encoding}
\State \emph{pattern specific decompression} \Comment {Mismatch byte load for zero
extend or partial match} \EndFor \State Store \emph{uncompressed words} \end{algorithmic} }
\end{algorithm}

\textbf{Compression.} Compressing data with C-Pack involves determining the
dictionary values that will be used to compress the rest of the line. In our
implementation, we serially add each word from the beginning of the cache
line to be a dictionary value if it was not already covered by a previous
dictionary value. For each dictionary value, we test whether the rest of the
words within the cache line is compressible. The next dictionary value is
determined using the predicate register to determine the next uncompressed
word, as in BDI. After four iterations (dictionary values), if all the
words within the line are not compressible, the cache line is left uncompressed.
Similar to BDI, the global predicate register is used to determine the
compressibility of all of the lanes after four or fewer iterations. Algorithm 6
provides the high-level algorithm for C-Pack compression. 

\begin{algorithm} \caption{C-PACK: Compression} \small{ \begin{algorithmic}[1]
\State load \emph{words} \For{each
dictionary value (including zero)} \Comment {To a maximum of four} \State \emph{test match/partial match}
\If{\emph{compressible}} \State Store \emph{encoding and mismatching
byte} \State
\textbf{break} \EndIf \EndFor \If{\emph{all lanes are compressible}} \State Store
\emph{compressed cache line} \EndIf \end{algorithmic} }
\end{algorithm} }

\subsection{Walkthrough of \SADA-based Compression} 
\begin{figure}[h] \centering \vspace{-0.0cm}
\includegraphics[width=0.42424242\textwidth]{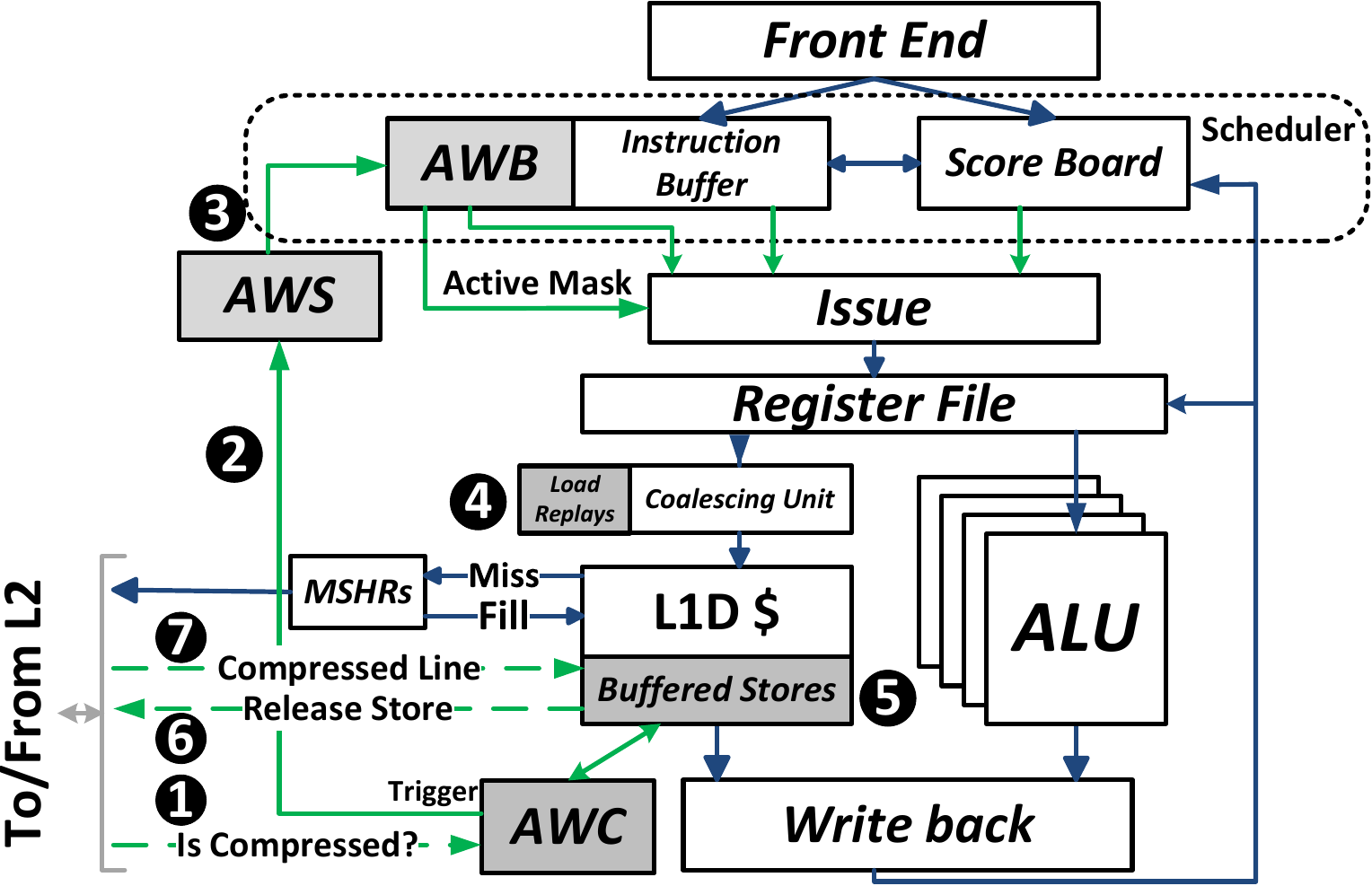}
\vspace{-0.06cm} \caption{Walkthrough of \SADA-based Compression. Figure
reproduced from \cite{caba}.}
\label{fig:load} \vspace{-0.1cm} \end{figure}

We show the detailed
operation of \SADA-based compression and decompression mechanisms in
Figure~\ref{fig:load}. 
%shows a walkthrough of the mechanism employed to enable
%decompression/compression of cache lines in the core.
We assume a baseline GPU architecture with three levels in the memory hierarchy
-- two levels of caches (private L1s and a shared L2) and main memory.
Different levels can potentially store compressed data. In this section and in
our evaluations, we assume that only the L2 cache and main memory contain
compressed data.\comm{\footnote{Different memory spaces exist in GPU
architectures.  In this work we apply compression only to global memory.}} Note
that \comm{even though the data is stored in the compressed form in both main
memory and L2 cache,} there is no capacity benefit in the baseline mechanism as
compressed cache lines still occupy the full uncompressed slot, i.e., we only
evaluate the bandwidth-saving benefits of compression in GPUs.

\subsubsection{The Decompression Mechanism.} Load instructions that access
global memory data in the compressed form trigger the appropriate \helperwarp to
decompress the data before it is used. The subroutines to decompress data are
stored in the \emph{Assist Warp Store (AWS)}. The AWS is indexed by the
compression encoding at the head of the cache line and by a bit indicating
whether the instruction is a load (decompression is required) or a store
(compression is required). Each decompression assist warp is given \emph{high
priority} and, hence, stalls the progress of its parent warp until it completes
its execution. This ensures that the parent warp correctly gets the decompressed
value. 

\comm{The AWS feeds into the AWB where it  available and active for scheduling.} 

\textbf{L1 Access.} We store data in L1 in the uncompressed form.  An L1 hit
does not require an assist warp for decompression.% execution flow.
 
\textbf{L2/Memory Access.} Global memory data cached in L2/DRAM could
potentially be compressed. A bit indicating whether the cache line is compressed
is returned to the core along with the cache line (\ding{202}). If the data is uncompressed,
the line is inserted into the L1 cache and the writeback phase resumes normally.
If the data is compressed, the compressed cache line is inserted into the L1
cache. The encoding of the compressed cache line and the warp ID
are relayed to the Assist Warp Controller (AWC), which then triggers the AWS
(\ding{203}) to deploy the appropriate \helperwarp (\ding{204}) to decompress the line. During regular execution, the load information for each thread is
buffered in the coalescing/load-store unit~\cite{coal1,coal2} until all the data
is fetched. We continue to buffer this load information (\ding{205}) until the
line is decompressed. 

%Normally load instructions are issued to the load/store pipeline where accesses
%from different threads are coalesced into fewer memory transactions. 
%
%In this case, however, the load instruction (to the compressed bytes) is not
%required to be processed and entries within the miss queue or free MSHRs are
%not required as it will always be an L1 hit. These checks and stages can be
%bypassed and the data can be directly loaded from the L1 caches (\ding{206})
%into the registers and written back after processing.
After the \SADA decompression subroutine ends execution, the original load that
triggered decompression is resumed (\ding{205}).

\comm{The registers are released at this point and regular execution is resumed
for the warp. }

\subsubsection{The Compression Mechanism.} \label{sec:comp} The \helperwarps to
perform compression are triggered by store instructions. When data is written to
a cache line (i.e., by a store), the cache line can be written back to main
memory either in the compressed or uncompressed form.  Compression is off the
critical path and the warps to perform compression can be scheduled when the
required resources are available. 

\comm{\begin{figure}[t] \centering \vspace{-0.4cm}
\includegraphics[width=0.42\textwidth]{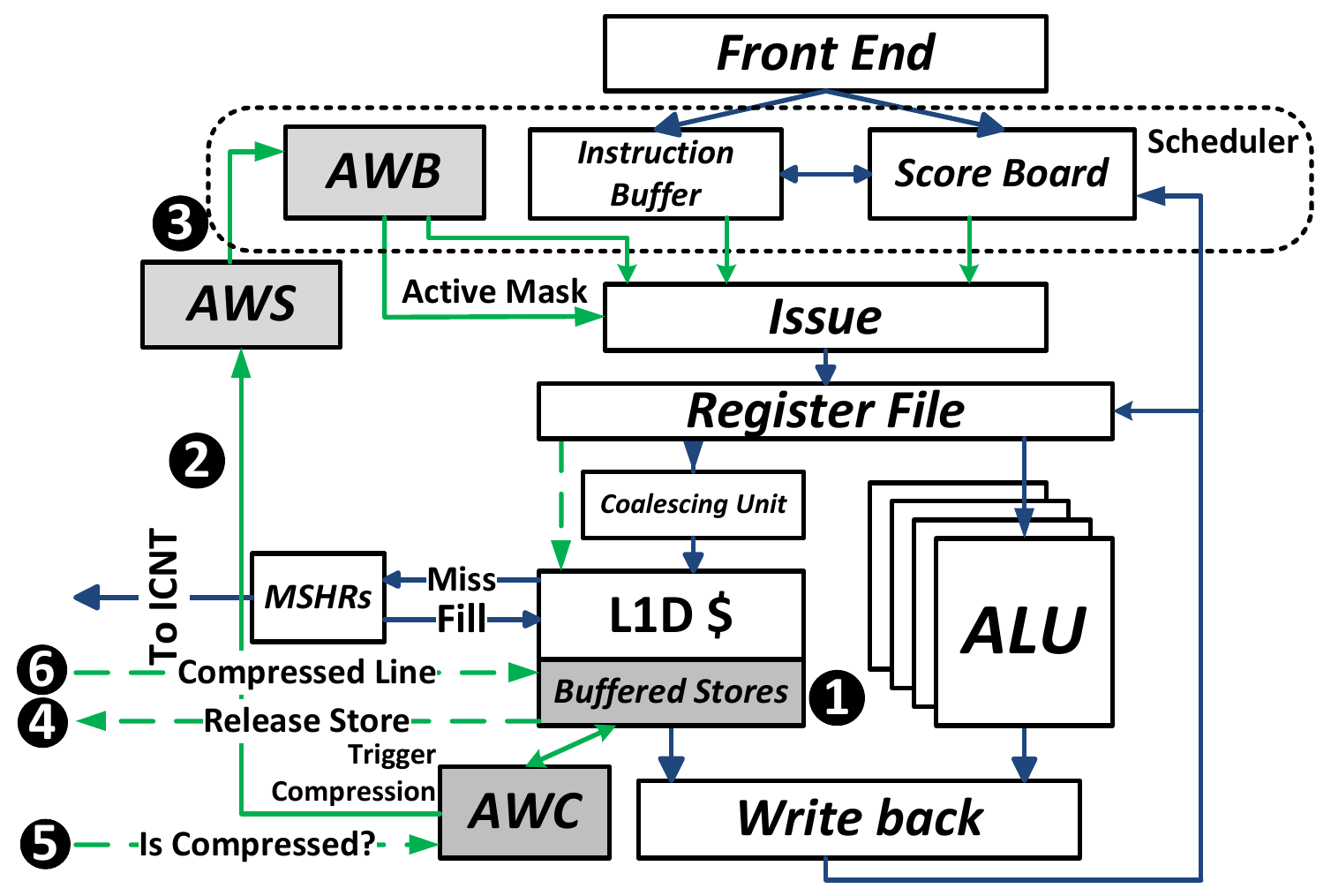} \vspace{-0.2cm}
\caption{Walkthrough of the Compression Mechanism. Figure reproduced from
\cite{caba}.} \label{fig:store}
\vspace{-0.2cm} \end{figure}}

\comm{The pending stores, however, need to be buffered until the corresponding
cache lines can be compressed. We consider two possible options for where to
buffer this data: (i) a special store buffer or (ii) a few dedicated sets within
the L1 cache (\ding{202}).\footnote{L1 caches are usually too small to fit the
working set of our applications, and hence the impact on performance of this
reduction in size is minimal.}}Pending stores are buffered in a few dedicated
sets within the L1 cache or in available shared memory (\ding{206}). In the case
of an overflow in this buffer space (\ding{206}), the stores are released to the
lower levels of the memory system in the uncompressed form (\ding{207}).  Upon
detecting the availability of resources to perform the data compression, the AWC
triggers the deployment of the \helperwarp that performs compression
(\ding{203}) into the AWB (\ding{204}), with \emph{low priority}. The scheduler
is then free to schedule the instructions from the compression subroutine. Since
compression is not on the critical path of execution, keeping such instructions
as low priority ensures that the main program is not unnecessarily delayed.  

\purple{\textbf{L1 Access.} On a hit in the L1 cache, the cache line is already
available in the uncompressed form. Depending on the availability of resources,
the cache line can be scheduled for compression or simply written to the L2 and
main memory uncompressed, when evicted.}  \comm{A write-evict policy is used in
the L1 (cite).}

\textbf{L2/Memory Access.} Data in memory is compressed at the granularity of a full cache line, but stores
can be at granularities smaller than the size of the cache line. This poses some
additional difficulty if the destination cache line for a store is already
compressed in main memory.  Partial writes into a compressed cache line would
require the cache line to be decompressed first, then updated with the new data,
and  written back to main memory. The common case{\textemdash}where the cache
line that is being written to is uncompressed initially{\textemdash}can be easily
handled. However, in the worst case, the cache line being partially written to
is already in the compressed form in memory. We now describe the mechanism to
handle both these cases.

Initially, to reduce the store latency, we assume that the cache line is uncompressed, and issue a store to the lower levels of the memory hierarchy, while buffering a copy in L1. If the cache line is found in L2/memory in the
uncompressed form (\ding{202}), the assumption was correct. The store then proceeds normally and the buffered stores are evicted from L1. If
the assumption is incorrect, the cache line
is
retrieved (\ding{208}) and decompressed before the store is retransmitted to the lower levels of the memory hierarchy.

\comm{If the cache line is found in L2/memory in the
uncompressed form (\ding{202}), the store proceeds normally. If the cache line
is already compressed, then it needs to be reloaded into the L1D (\ding{208})
and decompressed at the core before the store can be completed.  The store in
both these cases is buffered at the L1, but to reduce the store latency, is still transmitted to the lower
levels of the memory hierarchy assuming that the data is not compressed. If
this assumption is incorrect, we drop the store transaction and the cache line
is
retrieved, and then decompressed before the store is resumed. On the other
hand, if the cache line was correctly assumed to be uncompressed, a signal is
transmitted back to the core and the stores buffered at the L1 are
evicted.}

\subsection{Realizing Data Compression} Supporting data compression requires
additional support from the main memory controller and the runtime system, as we
describe below. 
 
\subsubsection{Initial Setup and Profiling.} Data compression with \SADA
requires a one-time data setup before the data is transferred to the GPU.
\comm{Typically, the input data that the GPU kernels operate on is initialized
by the CPU, stored in CPU main memory, and then transferred to the GPU DRAM by
invoking a memory copy function (e.g., \emph{``CUDAMemCopy''} in
CUDA~\cite{pguide}). The data is transferred to GPU DRAM across the PCI Express
link. }We assume initial software-based data preparation where the input data is
stored in CPU memory in the compressed form with an appropriate compression
algorithm before transferring the data to GPU memory. Transferring data in the
compressed form can also reduce PCIe bandwidth usage.\footnote{This requires
changes to the DMA engine to recognize compressed lines.} 

Memory-bandwidth-limited GPU applications are the best candidates for employing
data compression using \SADA. The compiler (or the runtime profiler) is required
to identify those applications that are most likely to benefit from this
framework. For applications where memory bandwidth is not a bottleneck, data
compression is simply disabled (this can be done statically or dynamically
within the framework).  \comm{Different kernels within an application
may also operate on different input data sets which may have varying levels of
compressibility. We use an initial sampling run to test the input data for
compressibility, and only compress those data sets which are most amenable. We
profile multiple different compression algorithms (e.g., BDI~\cite{bdi},
FPC~\cite{fpc}) to find the maximal opportunities for compression during the
profiling run.  }

\subsubsection{Memory Controller Changes.} \label{sec:mdcache} Data compression
reduces off-chip bandwidth requirements by transferring the same data in fewer
DRAM bursts. The memory controller (MC) needs to know whether the cache line
data is compressed and how many bursts (1--4 bursts in GDDR5~\cite{GDDR5}) are
needed to transfer the data from DRAM to the MC.  Similar to prior
work~\cite{GPUBandwidthCompression,lcp-micro}, we require metadata information
for every cache line that keeps track of how many bursts are needed to transfer
the data.\comm{For a maximum of 4 bursts, we need two bits of metadata for every
128B cache line.  Assuming a 4GB DRAM, we need $4GB / 128B * 2b = 8MB$ of
metadata space in DRAM.} Similar to prior work~\cite{GPUBandwidthCompression},
we simply reserve 8MB of GPU DRAM space for the metadata (\textasciitilde{0.2\%}
of all available memory).  Unfortunately, this simple design would require an
additional access for the metadata for every access to DRAM effectively doubling
the required bandwidth. To avoid this, a simple \emph{metadata (MD) cache} that
keeps frequently-accessed metadata on chip (near the MC) is required. Note that
this metadata cache is similar to other metadata storage and caches proposed for
various purposes in the memory controller,
e.g.,\cite{lcp-micro,page-overlays,meza-cal,qureshi-dram-caches,smart-refresh,raidr}. Our
experiments show that a small 8 KB 4-way associative MD cache is sufficient to
provide a hit rate of 85\% on average (more than 99\% for many applications)
across all applications in our workload pool.\footnote{For applications where MD
cache miss rate is low, we observe that MD cache misses are usually also TLB
misses. Hence, most of the overhead of MD cache misses in these applications is
outweighed by the cost of page table lookups.} Hence, in the common case, a
second access to DRAM to fetch compression-related metadata can be avoided.

\section{Methodology} \label{sec:methodology}

We model the \SADA framework in GPGPU-Sim 3.2.1~\cite{GPGPUSim}.
Table~\ref{tab:meth} provides the major parameters of the simulated system. We
use GPUWattch~\cite{gpuwattch} to model GPU power and CACTI~\cite{cacti} to
evaluate the power/energy overhead  associated with the MD cache
(Section~\ref{sec:mdcache}) and the additional components (AWS and AWC) of the CABA framework.  We implement BDI~\cite{bdi} using the Synopsys Design
Compiler with 65nm library (to evaluate the energy overhead of
compression/decompression for the dedicated hardware design for comparison to
\SADA), and then use ITRS projections~\cite{ITRS} to scale our results to the
32nm technology node.

\begin{table}[h!] 
	\vspace{-0.1cm} 
	\begin{scriptsize} 
	\centering
	\begin{tabular}{ll} 
\toprule System Overview           &  15 SMs, 32 threads/warp,  6 memory channels\\ 
\cmidrule(rl){1-2} Shader Core Config &  1.4GHz, GTO scheduler~\cite{tor-micro12}, 2 schedulers/SM\\
\cmidrule(rl){1-2} Resources / SM     &  48 warps/SM, 32768 registers, 32KB Shared Memory \\ 
\cmidrule(rl){1-2} L1 Cache    &  16KB, 4-way associative, LRU replacement policy    \\ 
\cmidrule(rl){1-2} L2 Cache   &  768KB, 16-way associative, LRU replacement policy  \\ 
\cmidrule(rl){1-2} Interconnect   &  1 crossbar/direction (15 SMs, 6 MCs), 1.4GHz  \\ 
\cmidrule(rl){1-2} Memory Model  &  177.4GB/s BW, 6 GDDR5 Memory Controllers (MCs),\\ 
		& FR-FCFS scheduling, 16 banks/MC\comm{, 924 MHz} \\ 
\cmidrule(rl){1-2}GDDR5 Timing~\cite{GDDR5}  & $t_{CL}=12,:t_{RP}=12,:t_{RC}=40,:t_{RAS}=28,$\\
		&$t_{RCD}=12,:t_{RRD}=6:t_{CLDR}=5:t_{WR}=12$ \\
%        \cmidrule(rl){1-2} Main Memory   &  2 partitions per controller (what does this mean?) 8? channels,FRFCFS,\\ & 8? banks-per-rank XX ranks per channel\\
  %      \cmidrule(rl){1-2} GDDR3 timing   &
  %      $(t_{CL}:t_{RP}:t_{RC}:t_{RAS}:t_{RCD}:t_{RRD}:t_{CLDR}:t_{WR}$ Fill in
  %      here~\cite{} \\ \cmidrule(rl){1-2}
%\multirow{2}[2]{*}{\centering Memory}              &  Timing: GDDR3
%(8-8-8)~\cite{micron} \\ & Organization: 1 channel, 1 rank-per-channel,\\ & 8
%banks-per-rank, 8 KB row-buffer \\
	\bottomrule \end{tabular}%
 \vspace{-0.1cm} \caption{Major parameters of the simulated systems.} 
\label{tab:meth}%
 \end{scriptsize}% 
\vspace{-0.4cm}
\end{table}%

\textbf{Evaluated Applications.} We use a number of CUDA applications derived
from CUDA SDK~\cite{sdk} (\emph{BFS, CONS, JPEG, LPS, MUM, RAY, SLA, TRA}),
Rodinia~\cite{rodinia} (\emph{hs, nw}), Mars~\cite{mars} (\emph{KM, MM, PVC, PVR, SS}) and
lonestar~\cite{lonestar} (\emph{bfs, bh, mst, sp, sssp}) suites.\comm{Note that
CUDA SDK implementation of breadth first search is different than that of
lonestar.}  We run all applications to completion or for 1 billion instructions
(whichever comes first).  \SADA-based data compression is beneficial mainly for
memory-bandwidth-limited applications.  In computation-resource limited applications,
data compression is not only unrewarding, but it can also cause significant
performance degradation due to the computational overheads associated with
assist warps. We rely on static profiling to identify memory-bandwidth-limited
applications and disable \SADA-based compression for the others.\comm{\footnote{The \SADA compression framework can
potentially adapt on-the-fly to different workloads and phases within a
workload.}} In our evaluation (Section~\ref{sec:Results}), we demonstrate
detailed results for applications that exhibit some compressibility in
memory bandwidth (at least 10\%). Applications without compressible data (e.g., sc, SCP) do not gain
any performance from the \SADA framework, and we verified that these
applications do not incur any performance degradation (because the assist warps
are \emph{not} triggered
for them).

\textbf{Evaluated Metrics.} We present Instruction per Cycle (\emph{IPC}) as the
primary performance metric. We also use \emph{average bandwidth
utilization}, defined as the fraction of total DRAM cycles that the DRAM data bus
is busy, and \emph{compression ratio}, defined as the ratio of the number of DRAM
bursts required to transfer data in the compressed vs. uncompressed form. As
reported in prior work~\cite{bdi}, we use decompression/compression latencies of 1/5
cycles for the hardware implementation of BDI.

\section{Results}
\label{sec:Results}
To evaluate the effectiveness of using \SADA to employ data compression, we
compare five different designs:
(i) \emph{Base} - the baseline system with no compression,
(ii) \emph{HW-BDI-Mem} - hardware-based \emph{memory bandwidth compression} with dedicated logic
(data is stored compressed in main memory but uncompressed in the last-level cache, similar to prior works~\cite{GPUBandwidthCompression,lcp-micro}),
(iii) \emph{HW-BDI} - hardware-based \emph{interconnect and memory bandwidth compression} 
(data is stored uncompressed only in the L1 cache)
(iv) \emph{CABA-BDI} - \SADAfull (\SADA) framework (Section~\ref{sec:idea}) with all associated overheads
of performing compression (for both interconnect and memory bandwidth),
(v) \emph{Ideal-BDI} - compression (for both interconnect and memory) with no
latency/power overheads for compression or decompression. This section provides our major results and analyses.\comm{All designs (unless stated otherwise)
  employ BDI compression algorithm~\cite{bdi}.}

%%% ONUR-final: If we have space, let's enumerate the other major results we provide in the technical report.

%the same as (ii) but with dedicated logic 
%to perform compression for on-chip interconnect traffic in addition to memory bandwidth,

\subsection{Effect on Performance and Bandwidth Utilization}

\begin{figure}[h!]
  \centering
   \vspace{-0.2cm}
  \includegraphics[width=0.49\textwidth]{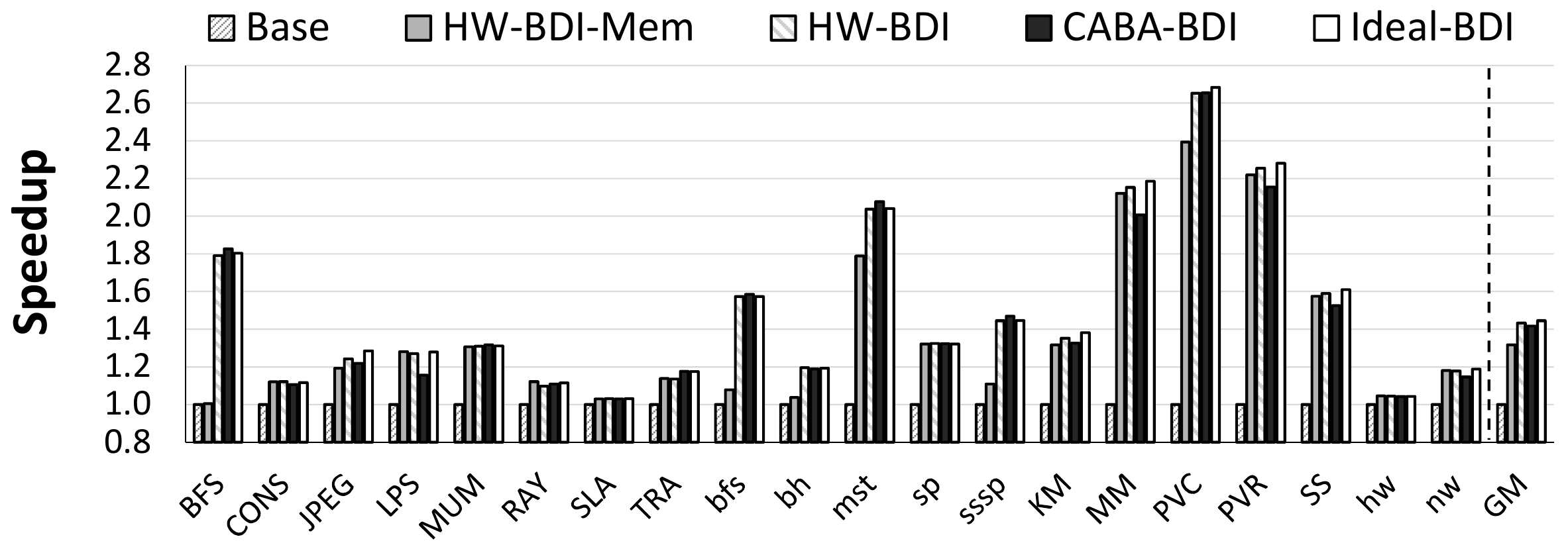}
   \vspace{-0.4cm}
  \caption{Normalized performance. Figure reproduced from ~\cite{caba}.}
  \label{fig:performance}
  \vspace{-0.2cm}
\end{figure}

\begin{figure}[h!]
  \centering
   \vspace{-0.4cm}
  \includegraphics[width=0.49\textwidth]{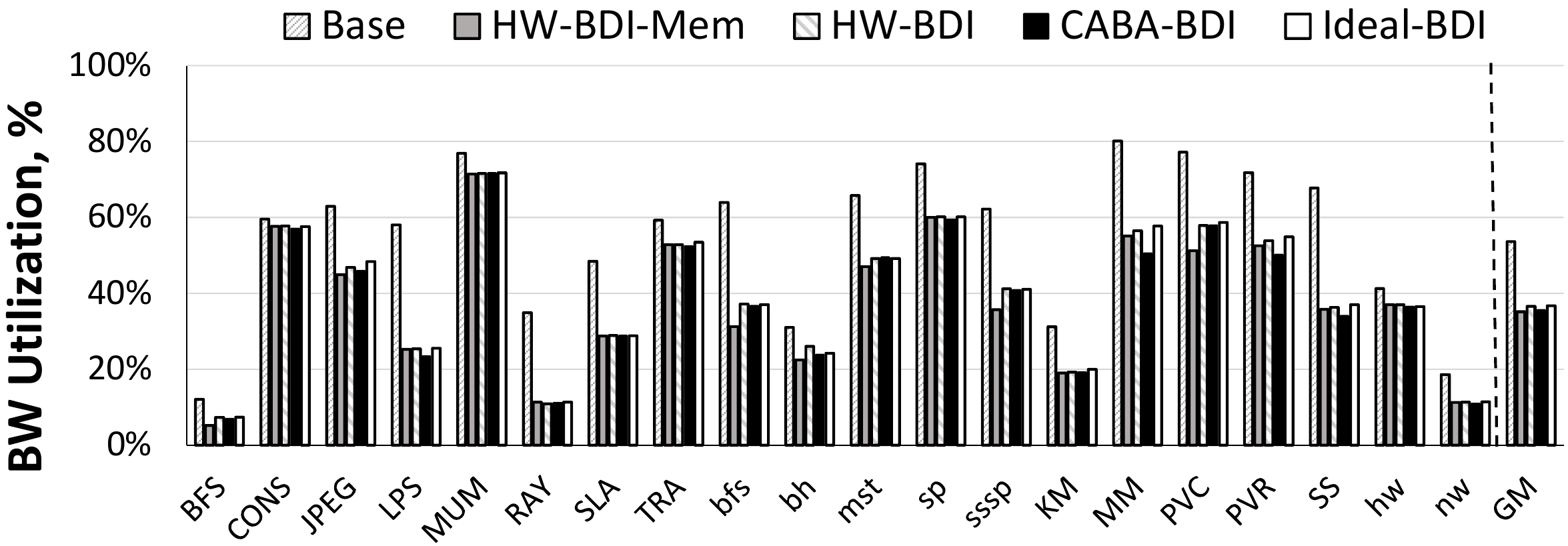}
   \vspace{-0.4cm}
  \caption{Memory bandwidth utilization. Figure reproduced from ~\cite{caba}.}
  \label{fig:bwutil}
  \vspace{-0.2cm}
\end{figure}

Figures~\ref{fig:performance} and~\ref{fig:bwutil} show, respectively,
the normalized performance (vs. \emph{Base}) and the memory bandwidth
utilization of the five designs.
%\footnote{
%All three compressed designs have very similar bandwidth utilization with the same compression
%algorithm. Hence we present results only for CABA-BDI.}
We make three major observations.

First, all compressed designs are effective in providing high
performance improvement over the baseline. Our approach (CABA-BDI)
provides a 41.7\% average improvement, which is only 2.8\% less than
the ideal case (Ideal-BDI) with none of the overheads associated with
\SADA. CABA-BDI's performance is 9.9\% better than the
previous~\cite{GPUBandwidthCompression} hardware-based memory
bandwidth compression design (HW-BDI-Mem), and \emph{only} 1.6\% worse
than the purely hardware-based design (HW-BDI) that performs both
interconnect and memory bandwidth compression. We conclude that our
framework is effective at enabling the benefits of compression without
requiring specialized hardware compression and decompression logic.

Second, performance benefits, in many workloads, correlate with the reduction in memory bandwidth utilization. For a fixed amount of data,
compression reduces the bandwidth utilization, and, thus, increases
the effective available bandwidth. Figure~\ref{fig:bwutil} shows that
CABA-based compression 1) reduces the average memory bandwidth
utilization from 53.6\% to 35.6\% and 2) is effective at alleviating the
memory bandwidth bottleneck in most workloads.\comm{Applications where bandwidth demand is high and where
  data is highly compressible with BDI compression algorithm (see
  Figure~\ref{fig:ratio}) are the ones that usually benefit most. For
  example, all applications from the Mars suite, \emph{bfs},
  \emph{mst}, and \emph{ssst} from the lonestar suite exhibit
  significant reduction in bandwidth utilization due to high data
  compressibility (up to 3.9X compression ratio for \emph{PVC}).
  This, in turn, results in high performance improvements for these
  applications (e.g., 2.6X performance improvement in \emph{PVC})} In
some applications (e.g., \emph{bfs} and \emph{mst}), designs that
compress \emph{both} the on-chip interconnect and the memory
bandwidth, i.e. CABA-BDI and HW-BDI, perform better than the design
that compresses only the memory bandwidth (HW-BDI-Mem). Hence, \SADA
seamlessly enables the mitigation of the interconnect bandwidth
bottleneck as well, since data compression/decompression is flexibly
performed at the cores.

Third, for some applications, CABA-BDI performs slightly (within 3\%) better
 than Ideal-BDI and HW-BDI. The reason for this
counter-intuitive result is the effect of warp
oversubscription~\cite{tor-micro12,nmnl-pact13,kayiran-micro14,medic}. In these cases, too
many warps execute in parallel, polluting the last level
cache. CABA-BDI sometimes reduces pollution as a side effect of
performing more computation in assist warps, which slows down the
progress of the parent warps.

%%% ONUR-final: ``normal warps'' or ``regular warps''? Let's pick one
%%% and be consistent everywhere in the paper. I am fine with either,
%%% but be consistent. Consistency rules.

We conclude that the \SADA framework can effectively enable data
compression to reduce both on-chip interconnect and off-chip memory
bandwidth utilization, thereby improving the performance of modern
GPGPU applications.

% (CABA-BDI) can significantly outperform previously proposed design
%(HW-BDI-Mem). The primary reason for this is that in our design the data moves in
%the compressed form not only through the off-chip bus, but also through the
%on-chip interconnect (this is why performance of HW-BDI is also better than
%HW-BDI-Mem for these applications). For the applications mentioned above, the on-chip
%interconnect bandwidth is another important bottleneck where data compression
%with CABA-BDI helps to improve performance. 
%For the same applications, CABA-BDI is
%slightly (within 3\%) better than Ideal-BDI. The reason for this
%counter-intuitive result is the effect of the warp
%oversubscription~\cite{tor-micro12,nmnl-pact13}. In these cases, too many warps
%are running in parallel, polluting the last level cache. CABA-BDI sometimes slows
%down this process as a side effect of adding additional code to the warps.
%
%We conclude that our proposed design (CABA-BDI) is an effective way of improving
%performance of modern GPGPU applications by reducing the bandwidth utilization
%of both off-chip and on-chip buses through data compression. 

\begin{comment}
\begin{figure}[h!]
  \centering
   \vspace{-0.4cm}
  \includegraphics[width=0.49\textwidth]{figures/CompRatioAll3.pdf}
   \vspace{-0.6cm}
  \caption{Compression ratio with BDI compression algorithm. Same for all compressed designs.}
  \label{fig:ratio}
  \vspace{-0.2cm}
\end{figure}
\end{comment}

\subsection{Effect on Energy}

\begin{figure}[h]
  \centering
   \vspace{-0.2cm}
  \includegraphics[width=0.49\textwidth]{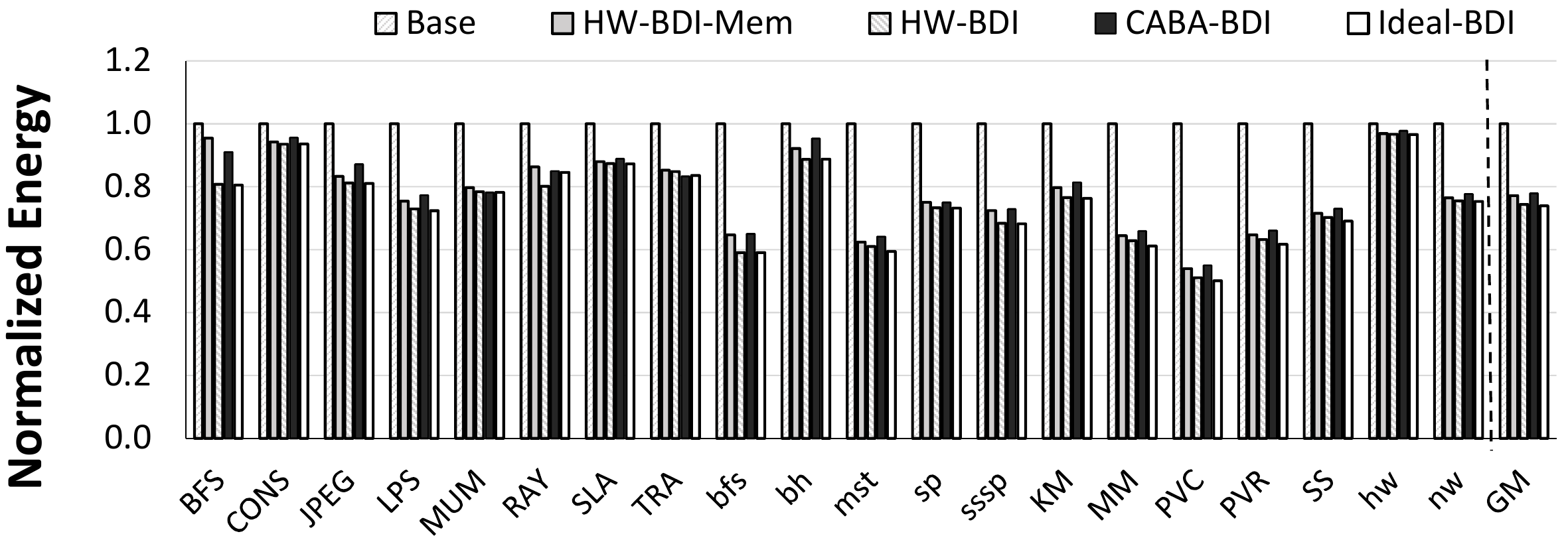}
   \vspace{-0.5cm}
  \caption{Normalized energy consumption. Figure reproduced from ~\cite{caba}.} 
  \label{fig:energy}
  \vspace{-0.2cm}
\end{figure}

Compression decreases energy consumption in two ways: 1) by reducing
bus energy consumption, 2) by reducing execution time.
Figure~\ref{fig:energy} shows the normalized energy consumption of the
five systems. We model the static and dynamic energy of the cores,
caches, DRAM, and all buses (both on-chip and off-chip), as well as
the energy overheads related to compression: metadata (MD) cache and
compression/decompression logic. We make two major observations.

%%% ONUR-final: Please fix all captions. 

First, CABA-BDI reduces energy consumption by as much as 22.2\% over
the baseline. This is especially noticeable for memory-bandwidth-limited
applications, e.g., \emph{PVC}, \emph{mst}.  This is a result of two
factors: (i) the reduction in the amount of data transferred between
the LLC and DRAM (as a result of which we observe a 29.5\% average reduction in DRAM power)
and (ii) the reduction in total execution time.  This observation
agrees with several prior works on bandwidth
compression~\cite{lcp-micro,memzip}.  We conclude that the \SADA
framework is capable of reducing the overall system energy, primarily
by decreasing the off-chip memory traffic.

Second, CABA-BDI's energy consumption is only 3.6\% more than that of
the HW-BDI design, which uses dedicated logic for memory bandwidth
compression. It is also only 4.0\% more than that of the Ideal-BDI
design, which has no compression-related overheads.\comm{The reduction
  in the execution time (due to less data sent over on-chip
  interconnect) usually alleviates the penalty of core-execution. It
  is also the primary cause for lower energy consumption with CABA-BDI
  in \emph{BFS} and \emph{RAY}.}  CABA-BDI consumes more energy
because it schedules and executes assist warps, utilizing on-chip
register files, memory and computation units, which is less
energy-efficient than using dedicated logic for compression. \purple{However, as
results indicate, this additional energy cost is small compared to the
performance gains of CABA (recall, 41.7\% over Base), and may be
amortized by using CABA for other purposes as well (see
Section~\ref{sec:Applications}).}

{\bf Power Consumption.} CABA-BDI increases the system power consumption by 2.9\%
over the baseline (not graphed), mainly due to the additional hardware
and higher utilization of the compute pipelines.\comm{However, the
  overheads incurred by the CABA-BDI mechanism are outweighed by the
  energy savings achieved for both off-chip and on-chip buses.}
However, the power overhead enables energy savings by reducing
bandwidth use and can be amortized across other uses of CABA
(Section~\ref{sec:Applications}).

\begin{figure}[h]
  \centering
   \vspace{-0.0cm}
  \includegraphics[width=0.49\textwidth]{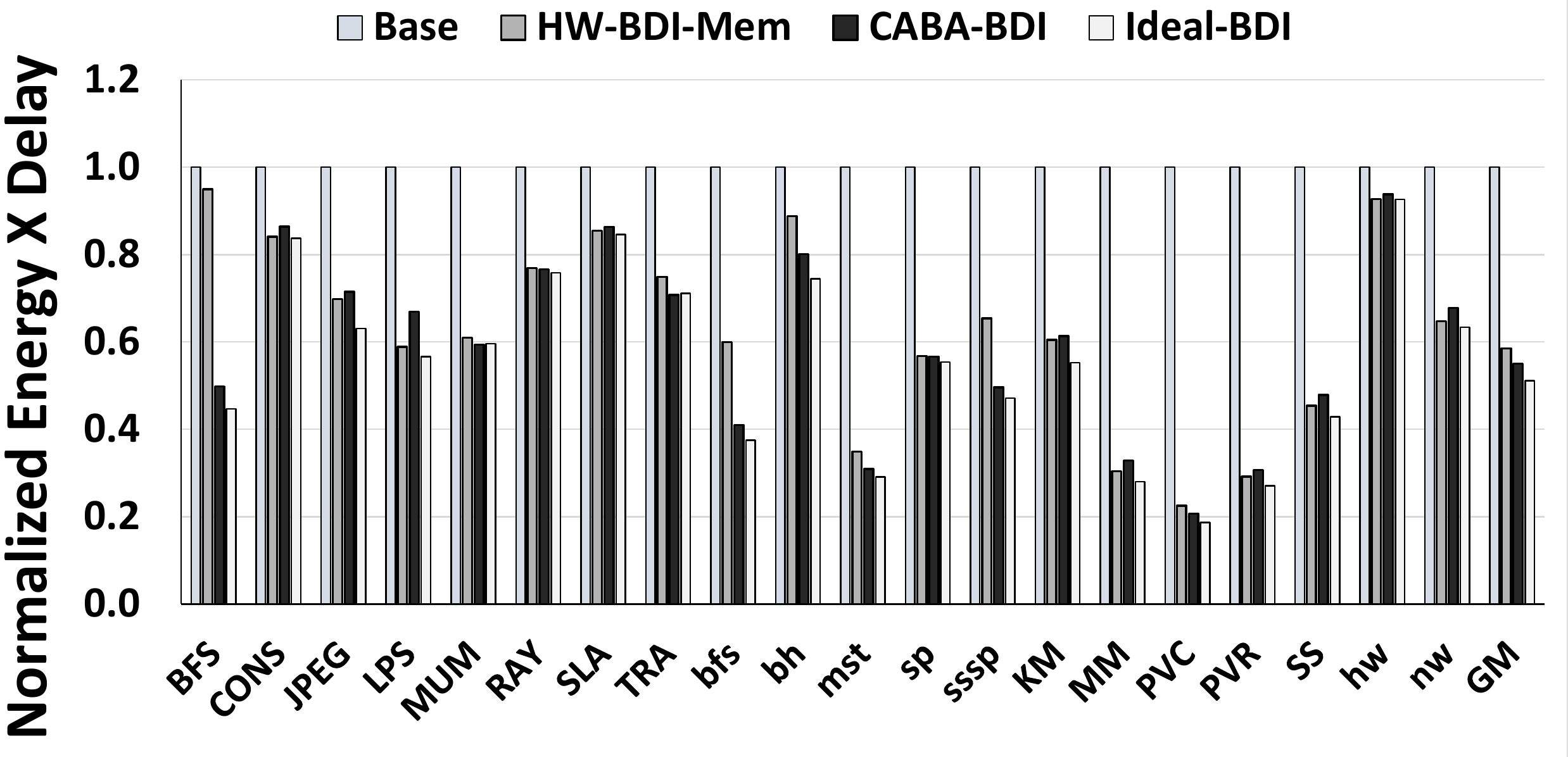}
  \caption{Energy-Delay product. }
  \label{fig:energy-perf}
  \vspace{-0.2cm}
\end{figure}

\blue{ {\bf Energy-Delay product.} Figure~\ref{fig:energy-perf} shows the product of the
normalized energy consumption and normalized execution time for the evaluated
GPU workloads. This metric simultaneously captures two metrics of
interest{\textemdash}energy dissipation and execution delay (inverse of
performance). An optimal feature
would simultaneously incur low energy overhead while also reducing the execution
delay. This metric is useful in capturing the efficiencies of different
architectural designs and features which
may expend differing amounts of energy while producing the same performance
speedup or vice-versa. Hence, a lower Energy-Delay product is more desirable. We observe that
CABA-BDI has a 45\% lower Energy-Delay product than the baseline. This reduction
comes from energy savings from reduced data transfers as well as lower execution
time. On average, CABA-BDI is within only 4\% of Ideal-BDI which incurs none of the
energy and performance overheads of the CABA framework. }

%Gena: remove it for now to save space

\subsection{Effect of Enabling Different Compression Algorithms}
The \SADA framework is \emph{not limited 
to a single compression algorithm}, and can be effectively used to employ other
hardware-based compression algorithms (e.g., FPC~\cite{fpc} and C-Pack~\cite{c-pack}).
The effectiveness of other algorithms depends on two key factors:
(i) how efficiently the algorithm maps to GPU instructions, (ii) how
compressible the data is with the algorithm. 
We map the FPC and C-Pack algorithms to the \SADA framework and evaluate the
framework's efficacy.\comm{Our technical report~\cite{caba-tr} details how
these algorithms are mapped to \SADA.} 
\comm{We estimate that FPC requires eight add instructions to compute the address 
of consecutive compressed words within a cache line, C-Pack -- X instructions.}
%, and hence has a higher 
%decompression overhead than BDI (BDI only requires one addition).

\begin{figure}[!h]
  \centering
   \vspace{-0.2cm}
  \includegraphics[width=0.49\textwidth]{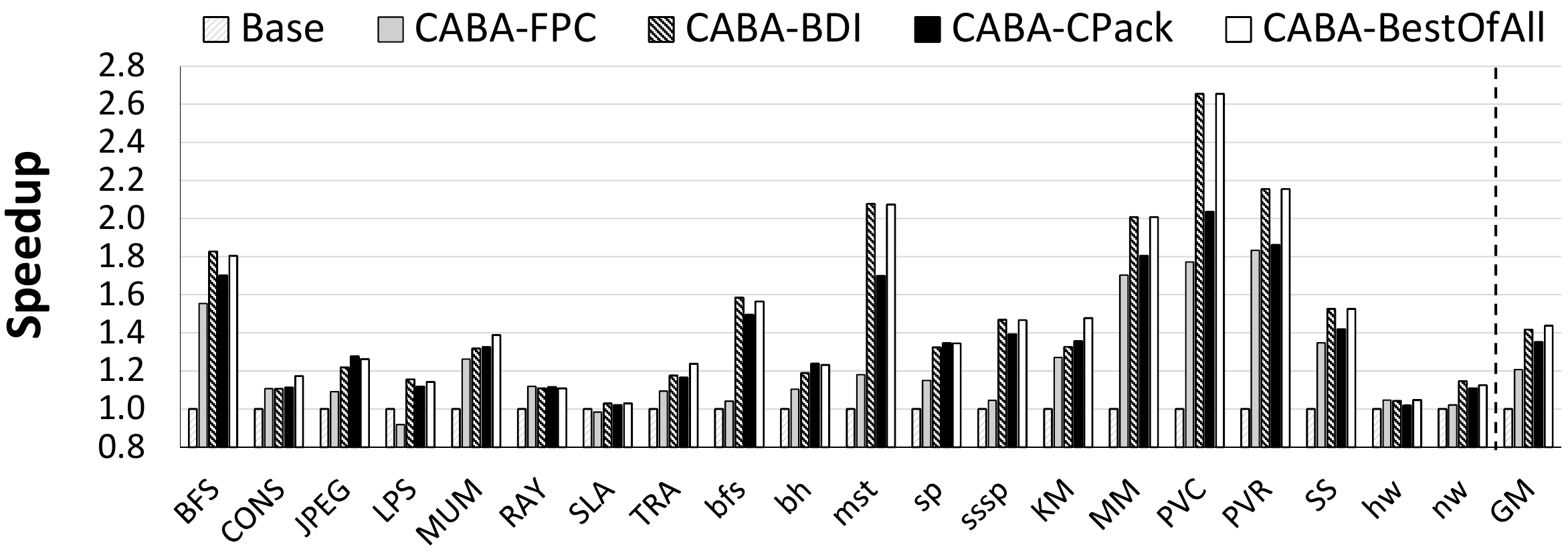}
   \vspace{-0.4cm}
  \caption{Speedup with different compression algorithms. Figure reproduced from
~\cite{caba}.}
  \label{fig:fpc}
  \vspace{-0.2cm}
\end{figure}

Figure~\ref{fig:fpc} shows the normalized speedup with four
versions of our design: \emph{CABA-FPC}, \emph{CABA-BDI},  \emph{CABA-C-Pack},
and \emph{CABA-BestOfAll}  
with the FPC, BDI, C-Pack compression algorithms. CABA-BestOfAll is an idealized design that selects and uses the best of all three algorithms in terms of compression ratio for \emph{each cache line}, assuming no selection overhead. 
We make three major observations.

\begin{figure}[h]
  \centering
   \vspace{-0.2cm}
  \includegraphics[width=0.49\textwidth]{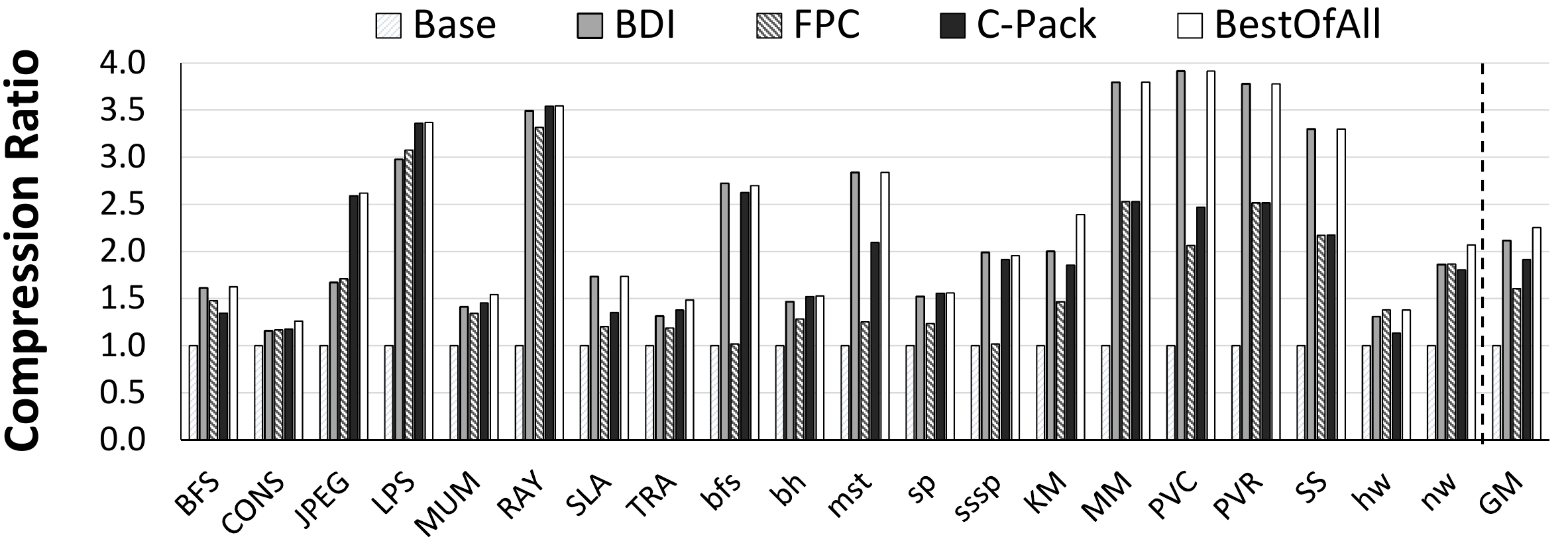}
   \vspace{-0.4cm}
  \caption{Compression ratio of algorithms with \SADA. Figure reproduced from
~\cite{caba}.}
  \label{fig:ratio}
  \vspace{-0.2cm}
\end{figure}

First, \SADA significantly improves performance with any compression algorithm (20.7\% with FPC, 35.2\% with C-Pack). Similar to CABA-BDI,
the applications that benefit the most are those that are both memory-bandwidth-sensitive (Figure~\ref{fig:bwutil})
and compressible (Figure~\ref{fig:ratio}). 
We conclude that our proposed framework, \SADA, is general and flexible enough
to successfully enable different compression
algorithms.

Second, applications benefit differently from each algorithm. 
For example, \emph{LPS, JPEG, MUM, nw} have higher compression ratios with FPC or C-Pack, 
whereas \emph{MM, PVC, PVR} compress better with BDI. 
This motivates the necessity of having \emph{flexible data compression} with different
algorithms within the same system. Implementing multiple compression
algorithms completely in hardware is expensive as it adds significant area
overhead, whereas \SADA can flexibly enable the use of different algorithms via
its general assist warp framework.

\comm{Third, the performance benefits with CABA-FPC are significantly lower than that of CABA-BDI.
There are two main reasons for this. First, as Figure~\ref{fig:ratio} shows, the compression
ratio with FPC is significantly lower than that of BDI
(2.1X vs. 1.6X correspondingly).\footnote{Our analysis shows that for our applications
the data patterns that can be compressed with BDI (e.g., array of pointers, pixels), but cannot
be compressed with FPC, are quite frequent. This leads to significantly higher compression ratios with BDI than
with FPC.} Second, the overhead of decompression with
FPC is higher than with BDI). This leads to a higher performance penalty when
applying CABA-FPC than CABA-BDI. In one application (\emph{LPS}), this overhead causes a degradation (e.g., 8.2\%)
in performance. In contrast, CABA-C-Pack, is much closer in performance to CABA-BDI, due to its competitive
compression ratio (1.9X vs. 2.1X with BDI).}

Third, the design with the best of three compression algorithms, CABA-BestOfAll,
can sometimes improve performance more than each
 individual design with just one compression algorithm (e.g., for \emph{MUM} and \emph{KM}). This 
happens because even within an application, different cache lines compress better with different algorithms. At the same time, different compression related overheads of different algorithms can cause one to have higher performance than another even though the latter may have a higher compression ratio. For example, CABA-BDI provides higher performance on \emph{LPS} than CABA-FPC, even though BDI has a lower compression ratio than FPC for \emph{LPS}, because BDI's compression/decompression latencies are much lower than FPC's. Hence, a mechanism
that selects the best compression algorithm based on \emph{both} compression ratio and the relative cost of compression/decompression is desirable to get the best of multiple compression algorithms.
The \SADA framework can flexibly enable the implementation of such a mechanism,
whose design we leave for future work. 

\subsection{Sensitivity to Peak Main Memory Bandwidth}
As described in Section~\ref{sec:motivation}, main memory (off-chip) bandwidth is a major
bottleneck in GPU applications. In order to confirm that \SADA works for different designs
with varying amounts of available memory bandwidth, we conduct an experiment where CABA-BDI is used in three systems with 0.5X, 1X and 2X amount of bandwidth
of the baseline.

\begin{figure}[h!]
  \centering
   \vspace{-0.17cm}
  \includegraphics[width=0.49\textwidth]{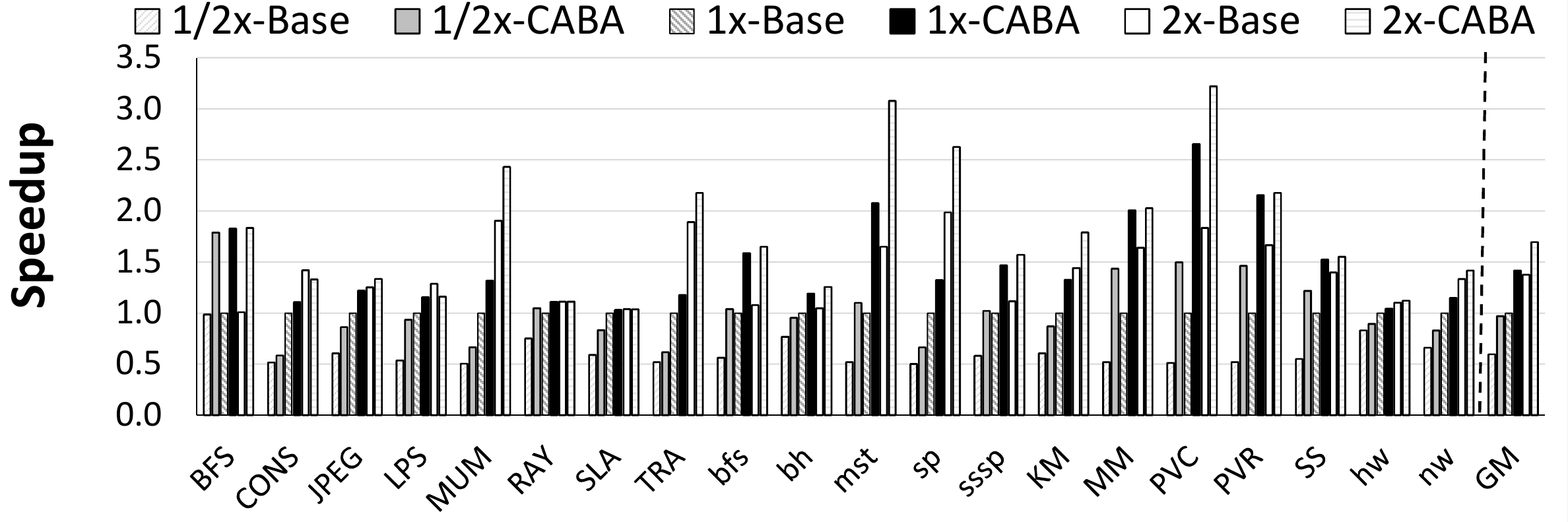}
   \vspace{-0.5cm}
  \caption{Sensitivity of \SADA to memory bandwidth. Figure reproduced from
~\cite{caba}.}
  \label{fig:bw}
  \vspace{-0.4cm}
\end{figure}

Figure~\ref{fig:bw} shows the results of this experiment. We observe that, as
expected, each \SADA design
(\emph{*-CABA}) significantly outperforms the corresponding baseline designs with the same
amount of bandwidth. The performance improvement of \SADA is often equivalent to
the doubling the off-chip memory bandwidth. We conclude that \SADA-based bandwidth compression, on average, offers almost
all the performance benefit of doubling the available off-chip
bandwidth with only modest complexity to support \helperwarps.  
 
\comm{
\subsubsection{Sensitivity to Decompression Latency}
Figure~\ref{fig:latency} shows the sensitivity of the performance to the decompression 
latency (varied from 0 in AW-Ideal to 16 instructions) assuming compression ratio achieved with BDI algorithm. 
As expected, the performance drops as we increase the latency of the decompression (that is on the critical
path of the execution). With 16 instructions for decompression the assist warp approach loses more than half of its
performance benefit with several applications degrading more than 20\% over the baseline. 
These results show some practical boundaries on how
complex can the decompression algorithm be (when it is mapped onto the GPU) to be efficient with our design.
 
\begin{figure}[h!]
  \centering
   \vspace{-0.2cm}
  \includegraphics[width=0.49\textwidth]{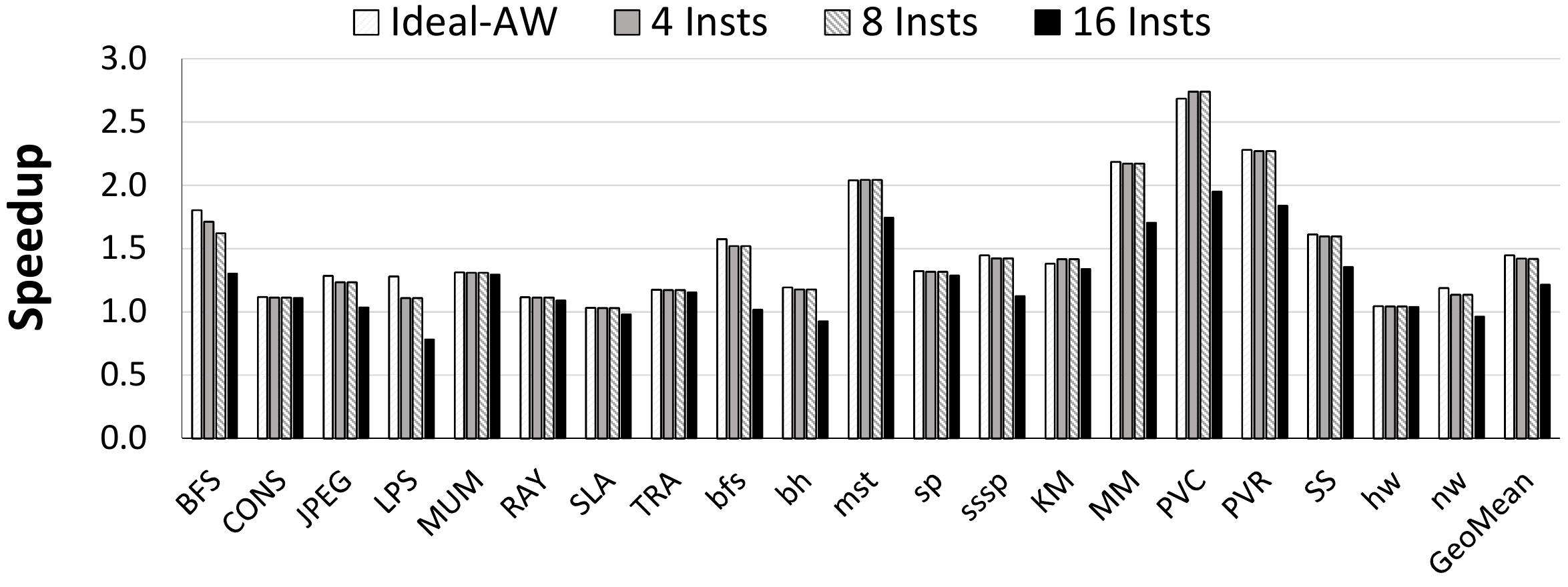}
   \vspace{-0.6cm}
  \caption{Effect on performance with different decompression latency: 0 (AW-Ideal), 4, 8, and 16 instructions. }
  \label{fig:latency}
  \vspace{-0.2cm}
\end{figure}
}
\comm{\subsubsection{Sensitivity to the Block/Atom Size}
\todo{Nandita}{This is an experiment where we vary cache line size: 64, 128, 256, and atom (chunk) size: 8,16,32}

\subsubsection{Sensitivity to Available Bandwidth}
Vary the bandwidth to see the effect on performance (and maybe energy).

\subsubsection{Sensitivity to Cache Size}
Vary L1 and L2 sizes to show that mostly bandwidth matters.

\subsubsection{Sensitivity to SM Number}
Vary the number of SMs to see the effect.}

%\subsection{\SADA Optimizations}
\subsection{Selective Cache Compression with \SADA}
\label{sec:caches}
 In addition to reducing bandwidth consumption, data compression can also
increase the \emph{effective capacity} of on-chip caches. While compressed caches can be
beneficial{\textemdash}as higher effective cache capacity leads to lower
miss rates{\textemdash}supporting cache compression requires several changes in the cache
design~\cite{fpc,c-pack,bdi,dcc,camp}. \comm{In this work, we do not aim to propose a new
compressed cache design, but investigate the capacity benefits of
compression in the L1/L2 caches based on the design originally proposed for
FPC~\cite{fpc}. }
%In general, we observe that cache compression can deliver up to
%X\% more performance for applications in our workload pool. 

\begin{figure}[!h] \centering \vspace{-0.0cm}
\includegraphics[width=0.49\textwidth]{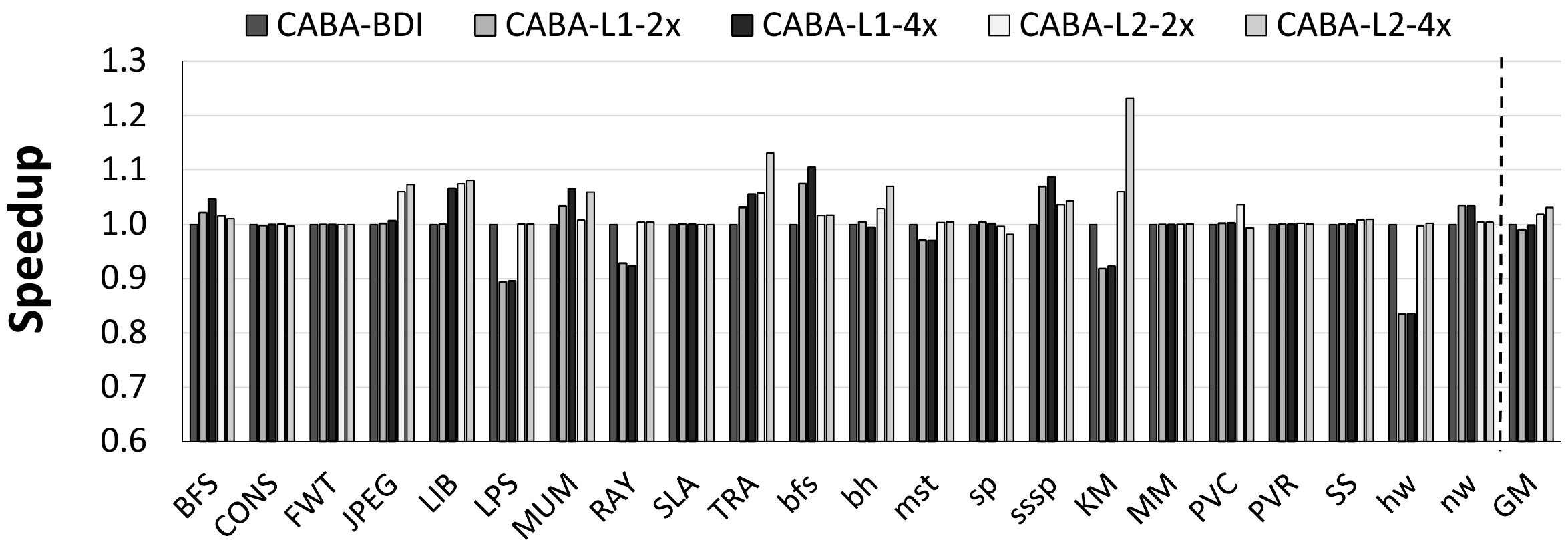}
 \vspace{-0.5cm}
\caption{Speedup of cache compression with \SADA.}
\label{fig:comp-caches} 
\vspace{-0.1cm} 
\end{figure}

 Figure~\ref{fig:comp-caches} shows the effect of four cache compression
designs using CABA-BDI (applied to both L1 and L2 caches
with 2x or 4x the number of tags of the baseline\footnote{The number of tags limits the effective compressed cache size~\cite{fpc,bdi}.}) on performance.  We make two major observations.
First, several applications from our workload pool are not only bandwidth
sensitive, but also cache capacity sensitive. For example, \emph{bfs} and \emph{sssp}
significantly benefit from L1 cache compression, while \emph{TRA} and
\emph{KM} benefit from L2 compression.  Second, L1 cache compression can severely
degrade the performance of some applications, e.g., \emph{hw} and \emph{LPS}.
The reason for this is the overhead of decompression, which can be
especially high for L1 caches as they are accessed very frequently.  This
overhead can be easily avoided by disabling compression at any level of the
memory hierarchy. 

\begin{figure}[!h] 
\centering 
\vspace{-0.2cm}
\includegraphics[width=0.49\textwidth]{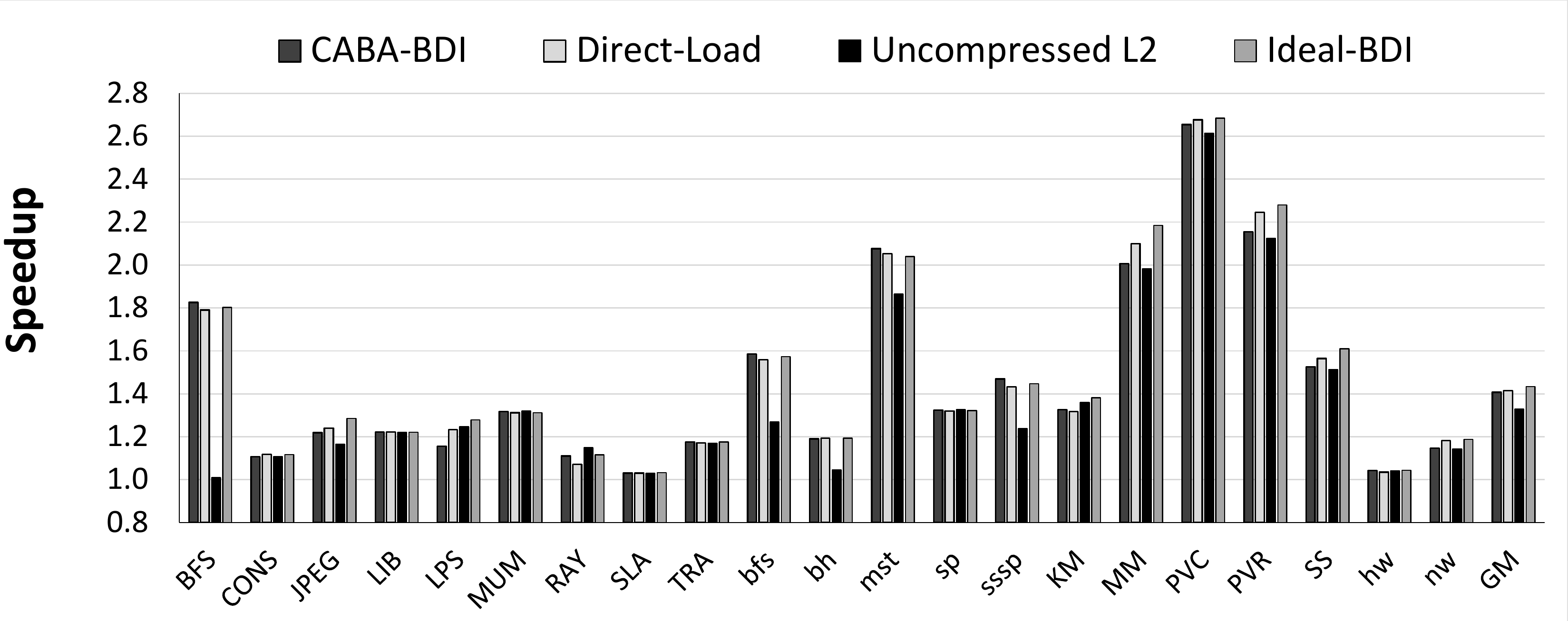}
 \vspace{-0.2cm}
\caption{Effect of different optimizations (Uncompressed data
in L2 and Direct Load) on applications' performance.} 
\label{fig:optimizations} 
\vspace{-0.2cm}
\end{figure}

\blue{\subsection{Other Optimizations}
We also consider several other optimizations of the \SADA framework for data
compression:
(i) avoiding
the overhead of decompression in L2 by storing data in the uncompressed form and 
(ii) optimized load of \emph{only useful} data.

\textbf{Uncompressed L2.}
\comm{This optimization provides another tradeoff between the savings in the on-chip traffic (when data in L2 is compressed -- default
option),
or savings in the additional latency of decompression (when data in L2 is uncompressed). 
Several applications in our workload pool (e.g., \emph{RAY}) benefit from
storing data uncompressed as these applications have higher hit rates in the L2
cache.}
The \SADA framework allows us to store compressed data selectively at different levels of the memory hierarchy. \comm{For example, \emph{either one or both}
levels of cache can hold uncompressed data with compression only at main memory
or \emph{all levels of memory} could hold compressed data. }We consider an
optimization where we avoid the overhead of decompressing data in L2 by storing
data in uncompressed form. This provides another
tradeoff between the savings in on-chip traffic (when data in L2 is
compressed -- default option), and savings in decompression latency (when data
in L2 is uncompressed). Figure~\ref{fig:optimizations} depicts the performance
benefits from this optimization. Several applications in our
workload pool (e.g., \emph{RAY}) benefit from storing data uncompressed as
these applications have high hit rates in the L2 cache. \purple{We conclude that
offering the choice of enabling or disabling compression at different levels of
the memory hierarchy can provide higher levels of the software stack (e.g.,
applications, compilers, runtime system, system software) with an additional performance knob.} \comm{\SADA framework provides application developers an additional
per-application performance knob with on-demand data compression
at different levels in the memory hierarchy.}

\textbf{Uncoalesced requests.} 
Accesses by scalar threads from the same
warp are coalesced into fewer memory transactions~\cite{pguide}. 
If the requests from different threads within a
warp span two or more cache lines, multiple lines have to be retrieved and
decompressed before the warp can proceed its execution. 
Uncoalesced requests can significantly increase the number of \helperwarps that need to be executed.
An alternative to decompressing each cache line (when only a few bytes from each line
may be required), is to enhance the coalescing unit to supply only the correct
\emph{deltas} from within each compressed cache line. 
The logic that maps bytes within a cache line to the appropriate registers will need to be
enhanced to take into account the encoding of the compressed line to determine
the size of the \emph{base} and the \emph{deltas}. 
As a result, we do not decompress the entire cache lines and only extract the
data that is needed. In this case, the cache line is not
inserted into the L1D cache in the uncompressed form, and hence every line needs to
be decompressed even if it is found in the L1D cache.\footnote{This
optimization also benefits cache lines that might \emph{not} have many uncoalesced
accesses, but have poor data reuse in the L1D.} \emph{Direct-Load} in
Figure~\ref{fig:optimizations} 
depicts the performance impact from this optimization. The overall
performance improvement is 2.5\% on average across all applications (as high as
4.6\% for \emph{MM}). }

%Figure~\ref{fig:optimizations} shows the effect of two potential optimizations

\section{Other Uses of the CABA Framework} \label{sec:Applications} The \SADA
framework can be employed in various ways to alleviate system bottlenecks and
increase system performance and energy efficiency. In this section, we discuss
two other potential applications of \SADA, focusing on two: \emph{Memoization} and
\emph{Prefetching}. We leave the detailed evaluations and analysis of these use
cases of CABA to future work. 

\subsection{Memoization} Hardware memoization is a technique used to avoid
redundant computations by reusing the results of previous computations that have
the same or similar inputs. Prior work~\cite{Arnau-memo, memoing2, sage}
observed redundancy in inputs to data in GPU workloads. In applications limited
by available compute resources, memoization offers an opportunity to trade off
computation for storage, thereby enabling potentially higher energy efficiency
and performance. In order to realize memoization in hardware, a look-up table
(LUT) is required to dynamically cache the results of computations as well as
the corresponding inputs. The granularity of computational reuse can be at the
level of fragments~\cite{Arnau-memo}, basic blocks or,
functions~\cite{alvarez-reuse,block-reuse,alvarez-reuse2,dynreuse,danconnors},
or long-latency instructions~\cite{memo-inst}.  The \SADA framework provides a natural
way to implement such an optimization. The availability of on-chip memory lends
itself for use as the LUT. In order to cache previous results in on-chip memory,
look-up tags (similar to those proposed in~\cite{unified-register}) are required
to index correct results. With applications tolerant of approximate results
(e.g., image processing, machine learning, fragment rendering kernels), the
computational inputs can be hashed to reduce the size of the LUT.  Register
values, texture/constant memory or global memory sections that are not subject
to change are potential inputs. \purple{An assist warp can be employed to
perform memoization in the following way: (1) compute the hashed value for
look-up at predefined trigger points, (2) use the load/store pipeline to save
these inputs in available shared memory, and (3) eliminate redundant
computations by loading the previously computed results in the case of a hit in
the LUT.}

\subsection{Prefetching} Prefetching has been explored in the context of
GPUs~\cite{Meng,osp-isca13,owl-asplos13,spareregister,kim,Arnau,Apogee} with the
goal of reducing effective memory latency. With memory-latency-bound
applications, the load/store pipelines can be employed by the \SADA framework to
perform opportunistic prefetching into GPU caches. The \SADA framework can
potentially enable the effective use of prefetching in GPUs due to several
reasons: (1) Even simple prefetchers such as the
stream~\cite{stream1,stream2,fdp} or stride~\cite{stride1,stride2} prefetchers
are non-trivial to implement in GPUs since access patterns need to be tracked
and trained at the fine granularity of warps~\cite{Apogee,kim}. CABA could enable
fine-grained book-keeping by using spare registers and assist warps to save
metadata for each warp. The computational units could then be used to
continuously compute strides in access patterns both within and across warps.
{(2) It has been demonstrated that software prefetching and helper
threads~\cite{ht2,ht5, ht7, \comm{ht10,}ht12, ht9, ht7,
ht1,spareregister,runahead-isca05, runahead-hpca03} are
very effective in performing prefetching for irregular access patterns. Assist
warps offer the hardware/software interface to implement application-specific
prefetching algorithms with varying degrees of complexity without the additional
overheads of various hardware implementations. (3) In bandwidth-constrained GPU systems,
uncontrolled prefetching could potentially flood the off-chip buses, delaying
demand requests. CABA can enable flexible prefetch throttling
(e.g.,~\cite{ebrahimi-micro09,ebrahimi-isca11,fdp}) by scheduling assist warps
that perform prefetching, \emph{only} when the memory pipelines are idle or underutilized.  (4)
Prefetching with CABA entails using load or prefetch instructions, which not
only enables prefetching to the hardware-managed caches, but also could simplify
the 
usage of unutilized shared memory or register file as prefetch buffers.

\blue{
\subsection{Other Uses} 

\textbf{Redundant Multithreading.} Reliability of GPUs is a key concern,
especially today when they are popularly employed in many supercomputing
systems. Ensuring hardware protection with dedicated resources can be
expensive~\cite{luo-dsn14}.
Redundant multithreading~\cite{gpu-rmt,rmt-alternatives,Qureshi-rmt,wang-rmt} is an approach where redundant threads are used to
replicate program execution. The results are compared at different points in
execution to detect and potentially correct errors. The CABA framework can be
extended to redundantly execute portions of the original
program via the use of such approaches to increase the reliability
of GPU architectures.

\textbf{Speculative Precomputation.} In CPUs, speculative
multithreading (~\cite{ht16, ht17, ht18}) has
been proposed to speculatively parallelize serial code and verify the
correctness later. Assist
warps can be employed in GPU architectures to speculatively
pre-execute sections of code during idle cycles to further improve parallelism
in the program execution. Applications tolerant to approximate results could
particularly be amenable towards this optimization~\cite{RFVP}.

\textbf{Handling Interrupts and Exceptions.} Current GPUs do not implement support for
interrupt handling except for some support for timer interrupts used
for application time-slicing~\cite{fermi}. CABA offers a natural mechanism for
associating architectural events with subroutines to be executed in
throughput-oriented architectures where thousands of threads could be active at
any given time. Interrupts and exceptions can be handled by special assist
warps, without requiring complex context switching or heavy-weight kernel
support. 
  
\textbf{Profiling and Instrumentation.} 
Profiling and binary instrumentation tools like Pin~\cite{pin} and Valgrind~\cite{valgrind}
proved to be very useful for development, performance analysis and debugging
on modern CPU systems. At the same time, there is a lack~\footnote{With the 
exception of one recent work~\cite{gpuPin}.} of tools with
same/similar capabilities for modern GPUs. This significantly limits software development
and debugging for modern GPU systems. The CABA framework can potentially enable easy and efficient
development of such tools, as it is flexible enough to invoke user-defined
code on specific architectural
events (e.g., cache misses, control divergence). 
}

\comm{\todo{Nandita}{Write up a few lines on each below:}
\begin{itemize} \item Redundant multithreading for reliability \item Speculative
precomputation \item Dynamic data-movement/re-ordering for better locality \item
Profiling and code instrumentation (thanks Gena) \end{itemize} }

\comm{-As we previously described in Section~\ref{sec:background}, data
compression techniques can be an effective way -to relax bandwidth demands of
the modern GPUs. Similar to prior works (e.g., \cite{GPUBandwidthCompression})
-we expect the benefits from bandwidth compression by exploiting the possibility
of sending -variable number of GPU DRAM bursts~\cite{GDDR5}. Unfortunately,
prior work usually rely on the hardware-based -implementation of both
compression and decompression, that requires separate hardware design and
implementation.
	-
	-A SADA-based framework, in contrast, can utilize the existing GPU
pipelines to perform compression/decompression -on demand (when bandwidth is
limited), and hence avoid additional complexity and cost of the dedicated
compression -algorithm implementation. It generates the \helperwarps to perform
decompression/compression -that are triggered by load/store instructions that
access global memory.  -The sequence of instructions that are inserted into the
active \helperwarp buffer depend on the type of load/store -instruction as well
as corresponding data. The primary advantage of a SADA-based framework for data
compression -is its ability to utilize existing GPU resources that are otherwise
can be significantly underutilized. }

%\vspace{-0.1cm}
\section{Related Work} \label{sec:Related} 

To our knowledge, this paper is the first to (1) propose a flexible and
general framework for employing idle GPU resources for useful
computation that can aid regular program execution, and (2) use the
general concept of {\em helper threading} to perform memory and
interconnect bandwidth compression. We demonstrate the benefits of our
new framework by using it to implement multiple compression algorithms
on a throughput-oriented GPU architecture to alleviate the
memory bandwidth bottleneck. In this section, we discuss related
works in helper threading, and memory bandwidth optimizations, and memory
compression.

\textbf{Helper Threading.} Previous
  works~\cite{ssmt,ssmt2,ht2,ht5\comm{,ht0},ht20,ht12,ht0,assisted-execution,
    assisted-execution-04,ht11,ht19,ht25,ht4,ht7,ht9,ht1,runahead-hpca03,runahead-isca05,zilles-exception} demonstrated
  the use of \emph{helper threads} in the context of Simultaneous
  Multithreading (SMT) and multi-core and single-core processors, primarily to speed
  up single-thread execution by using idle SMT contexts, idle
  cores in CPUs, or idle cycles during which the main program is stalled on a
single thread context. These works typically use helper threads (generated
  by the software, the hardware, or cooperatively) to pre-compute
  useful information that aids the execution of the primary thread
  (e.g., by prefetching, branch outcome pre-computation, and cache
  management). No previous work discussed the use of helper threads
  for memory/interconnect bandwidth compression or cache compression.

These works primarily use helper threads to capture data flow and pre-compute useful information to aid in the execution of the primary thread. In these prior works, helper threads are either
generated with the help of the compiler~\cite{ht11, ht19, ht25} or completely
in hardware~\cite{ht4, runahead-isca05, runahead-hpca03}. These threads are used to perform prefetching in
~\cite{ssmt, ht2,ht5, ht7, ht10, ht12, ht9, ht7, ht1, runahead-hpca03,
runahead-isca05} where the helper threads predict future load addresses by doing some computation and then
prefetch the corresponding data. Simultaneous Subordinate Multithreading ~\cite{ssmt} employs hardware generated
helper threads to improve branch prediction accuracy and cache management. Speculative
multi-threading~\cite{ht15, ht16,ht17, ht18} involves executing different
sections of the serial program in parallel and then later verifying the
run-time correctness. Assisted execution~\cite{assisted-execution,
assisted-execution-04} is an execution paradigm where tightly-coupled
\emph{nanothreads} are generated using \emph{nanotrap handlers} and execute
routines to enable optimizations like prefetching.
In Slipstream Processors~\cite{ht7}, one thread runs ahead of the program
and executes a reduced version of the program. In runahead
execution~\cite{runahead-hpca03,runahead-isca05}, the main thread is executed
speculatively solely for prefetching purposes when the program is stalled due to a cache
miss. 

While our work was inspired by these prior studies of helper threading
in latency-oriented architectures (CPUs), developing a framework for
helper threading (or {\em assist warps}) in throughput-oriented
architectures (GPUs) enables new opportunities and poses new
challenges, both due to the massive parallelism and resources present
in a throughput-oriented architecture (as discussed in
Section~\ref{sec:intro}).  Our CABA framework exploits these new
opportunities and addresses these new challenges, including (1)
low-cost management of a large number of assist warps that could be running
concurrently with regular program warps, (2) means of state/context
management and scheduling for assist warps to maximize effectiveness
and minimize interference, and (3) different possible applications of
the concept of assist warps in a throughput-oriented architecture.

\comm{
%%% ONUR-final: We really do not need these. Is there any reason why
%%% the following works were not cited above when you discussed helper
%%% threads?
%%% Besides, some references are duplicated below, e.g., 42...
%%% Finally, speculative multithreading is really not relevant. We need space, so let's not derail related work.
In these prior works, helper threads are either generated with the
help of the compiler (e.g.,~\cite{ht11, ht19, ht25}) or completely in
hardware (e.g.,~\cite{ht4}) and used to perform optimizations like
prefetching in ~\cite{ssmt, ht2,ht5, ht7, ht10, ht12, ht9, ht7, ht1,
  assisted-execution-04}, branch prediction~\cite{ssmt}, cache
management~\cite{ssmt} or speculative
multi-threading~\cite{ht15,ht16,ht17,ht18}.
}

In the GPU domain, CudaDMA~\cite{cudadma} is a recent proposal that
aims to ease programmability by decoupling execution and memory
transfers with specialized DMA warps. This work does \emph{not} provide a
general and flexible hardware-based framework for using GPU cores to run warps that aid the main program.

% and improve memory bandwidth
%  utilization by exploiting memory-level parallelism more
%  efficiently.}

 \comm{In Slipstream Processors~\cite{ht7}, one thread runs ahead of the program
and executes a reduced version of the program.}

\blue{\textbf{Memory Latency and Bandwidth Optimizations in GPUs.}
  A lot of prior works focus on optimizing for memory bandwidth and memory
latency in GPUs.
  \comm{Lakshminarayana et al.~\cite{spareregister} and }Jog et
  al.~\cite{osp-isca13} aim to improve memory latency tolerance by coordinating
prefetching and warp scheduling policies. Lakshminarayana et al.~\cite{spareregister} reduce effective latency in graph applications by using spare registers to store prefetched data. In OWL~\cite{owl-asplos13}
  and~\cite{nmnl-pact13}, intelligent scheduling is used to improve
  DRAM bank-level parallelism and bandwidth utilization and Rhu et
  al.~\cite{rhu-micro13} propose a locality-aware memory to improve
  memory throughput. Kayiran et al.~\cite{nmnl-pact13} propose GPU throttling techniques to reduce
memory contention in heterogeneous systems. Ausavarangniran et al.~\cite{medic}
leverage heterogeneity in warp behavior to design more intelligent policies at
the cache and memory controller. These works do not consider data compression and
  are orthogonal to our proposed framework. }

\textbf{Compression.} Several prior
works~\cite{LinkCompression,CompressionPrefetching,GPUBandwidthCompression,
  lcp-micro,memzip, sc2, camp,toggle-aware-cal,toggle-aware-hpca} study memory and cache compression with
several different compression
algorithms~\cite{fpc,bdi,c-pack,sc2,zvc,fvc}, in the context of CPUs
or GPUs. 

\blue{Alameldeen et al.~\cite{CompressionPrefetching} investigated the possibility of
bandwidth compression with FPC~\cite{fpc-tr}. The authors show that
significant decrease in pin bandwidth demand can be achieved with FPC-based
bandwidth compression design. Sathish et al.~\cite{GPUBandwidthCompression}
examine the GPU-oriented memory
link compression using C-Pack~\cite{c-pack} compression algorithm. The authors make
the observation that GPU memory (GDDR3~\cite{gddr3}) indeed allows transfer of data
in small bursts and propose to store data in compressed form in memory,
but without capacity benefits. Thuresson et al.~\cite{LinkCompression} consider a CPU-oriented design where
a compressor/decompressor logic is located on the both ends of the main memory
link. Pekhimenko et al.~\cite{lcp-micro} propose
Linearly Compressed Pages (LCP) with the primary goal of compressing main memory
to increase capacity.}

Our work is the first to demonstrate how one can adapt some
of these algorithms for use in a general helper threading framework
for GPUs. As such, compression/decompression using our new framework
is more flexible since it does not require a specialized hardware
implementation for any algorithm and instead utilizes the existing GPU
core resources to perform compression and decompression. Finally, as
discussed in Section~\ref{sec:Applications}, our CABA framework is
applicable beyond compression and can be used for other purposes.

%Thuresson et al.~\cite{LinkCompression} considered a CPU-oriented design where
%a compressor/decompressor logic is located on the both ends of the main memory
%link, and, hence, every data transfer between the main memory and the last
%level cache requires both compression and decompression.  This design assumed
%an abstract link model, that does not take into account certain limitations of
%the modern main memory technologies, e.g., DRAM. This paper has no discussion
%on how variable blocks can be trasferred in the context of modern CPU
%DRAMs~\cite{DDR3}. 

%In the proposed design the bandwidth bottleneck is now between main memory and
%compressor/decompressor, and in order to keep up with the speed of the main
%memory link, you need either to increase the latency or the width of the
%connection between the main memory and the compressor/decompressor.  This is
%difficult to do with the modern DRAM design due to both (i) thermal and power
%issues and (ii) complexity.

\comm{Alameldeen et al.~\cite{CompressionPrefetching} investigated the possibility of
bandwidth compression with FPC~\cite{fpc-tr}.  Authors showed that significant
decrease in pin bandwidth demand can be achieved with FPC-based bandwidth
compression design, but assumed variable-size data transfers that is not
possible with modern DRAMs~\cite{ddr3-jedec}. Sathish et
al.~\cite{GPUBandwidthCompression} look at the GPU-oriented memory link
compression using C-Pack~\cite{c-pack} compression algorithm. Authors make the
observation that GPU memory (GDDR3~\cite{gddr3}) indeed allows transfer of data
in small bursts and propose to store data in the compressed form in the memory,
but without space benefits.  Unfortunately, this work still have two major
drawbacks mentioned previously. Pekhimenko et al.~\cite{lcp-micro} proposed
Linearly Compressed Pages (LCP) with the primary goal of compressing main memory
to increase capacity. This design still relies on all the algorithms being
implemented in hardware.  \comm{This design also has several drawbacks.  First,
bandwidth savings are achieved by bringing additional adjacent cache lines that
fit in a single memory transfer that only works when applications exhibit
spatial locality (and can potentially result in a cache pollution).  Second,
even though LCP design can work with multiple t compression algorithms, it still
relies on all the algorithms being implemented in hardware.}}

%\vspace{-0.1cm}
\section{Conclusion} 

This paper makes a case for the \SADAfull (\SADA) framework, which
employs \emph{assist warps} to alleviate different
bottlenecks in GPU execution. \SADA is based on the key observation
that various imbalances and bottlenecks in GPU execution leave on-chip
resources, i.e., computational units, register files and on-chip
memory, underutilized. CABA takes advantage of these idle resources and employs
them to perform useful work that can aid the execution of the main program and
the system. 

We provide a detailed design and analysis of
how \SADA can be used to perform flexible data compression in GPUs to
mitigate the memory bandwidth bottleneck.  Our extensive evaluations
across a variety of workloads and system configurations show that the
use of \SADA for memory compression significantly improves system
performance (by 41.7\% on average on a set of bandwidth-sensitive GPU
applications) by reducing the memory bandwidth requirements of both the
on-chip and off-chip buses.\comm{We also show that \SADA can flexibly
implement multiple different compression algorithms, which have
different effectiveness for different applications or data
patterns.} 

We conclude that CABA is a general substrate that can
alleviate the memory bandwidth bottleneck in modern GPU systems by
enabling flexible implementations of data compression algorithms.  We
believe \SADA is a general framework that can have a wide set of use
cases to mitigate many different system bottlenecks in
throughput-oriented architectures, and we hope that future work
explores both new uses of \SADA and more efficient implementations of
it.

%be used for
%  various other applications to mitigate other system bottlenecks, and
%  we hope future work explores new uses of \SADA.}

% applications are not limited to bandwidth,
%and it can be potentially applied to mitigate other system bottlenecks (e.g.,
%limited available computational units bottleneck can be alleviated by performing
%memoization).
%
%Our extensive evaluations across a variety of workloads and system
%configurations show that Core-Assisted Data Compression can significantly
%improve system performance (41.7\% on average on a subset of bandwidth sensitive
%applications) by reducing bandwidth requirements of both the on-chip and
%off-chip buses.  The average performance improvement is close to the baseline
%design where the off-chip bandwidth is doubled. We conclude that \SADA is an
%efficient substrate to alleviate the memory bandwidth bottleneck using data
%compression in modern GPU systems.

%\vspace{-0.1cm}
\section*{Acknowledgments}

We thank the reviewers for their valuable suggestions. We thank the
members of the SAFARI group for their feedback and the stimulating research
environment they provide. Special thanks to Evgeny Bolotin and Kevin Hsieh for their feedback during
various stages of this project. We acknowledge the support of our
industrial partners: Facebook, Google, IBM, Intel, Microsoft, Nvidia, 
Qualcomm, VMware, and Samsung. This research was partially supported
by NSF (grants 0953246, 1065112, 1205618, 1212962, 1213052, 1302225,
1302557, 1317560, 1320478, 1320531, 1409095, 1409723, 1423172,
1439021, 1439057), the Intel Science and Technology Center for
Cloud Computing, and the Semiconductor Research Corporation.  Gennady Pekhimenko is supported in part by a Microsoft
Research Fellowship and an Nvidia Graduate Fellowship. Rachata Ausavarungnirun
is supported in part by the Royal Thai Government scholarship. This article is a
revised and extended version of our previous ISCA 2015 paper~\cite{caba}. 

\pagebreak

%\bstctlcite{bstctl:etal, bstctl:nodash, bstctl:simpurl}
\begin{spacing}{0.85}
\begin{footnotesize}
\bibliographystyle{plain}
\bibliography{paper}

\begin{thebibliography}{100}

\bibitem{ht1}
T.~M. Aamodt et~al.
\newblock Hardware support for prescient instruction prefetch.
\newblock In {\em HPCA}, 2004.

\bibitem{MXT}
B.~Abali et~al.
\newblock {Memory Expansion Technology ({MXT}): Software Support and
  Performance}.
\newblock {\em {IBM J.R.D.}}, 2001.

\bibitem{warped-register}
M.~Abdel-Majeed et~al.
\newblock Warped register file: A power efficient register file for gpgpus.
\newblock In {\em HPCA}, 2013.

\bibitem{fpc}
A.~Alameldeen et~al.
\newblock {Adaptive Cache Compression for High-Performance Processors}.
\newblock In {\em ISCA}, 2004.

\bibitem{fpc-tr}
A.~Alameldeen et~al.
\newblock {Frequent Pattern Compression: A Significance-Based Compression
  Scheme for {L2} Caches}.
\newblock Technical report, {U. Wisconsin}, 2004.

\bibitem{CompressionPrefetching}
A.~Alameldeen et~al.
\newblock Interactions between compression and prefetching in chip
  multiprocessors.
\newblock In {\em HPCA}, 2007.

\bibitem{alvarez-reuse}
C.~Alvarez et~al.
\newblock On the potential of tolerant region reuse for multimedia
  applications.
\newblock In {\em ICS}, 2001.

\bibitem{memoing2}
C.~Alvarez et~al.
\newblock Fuzzy memoization for floating-point multimedia applications.
\newblock {\em IEEE Trans. Comput.}, 2005.

\bibitem{alvarez-reuse2}
C.~Alvarez et~al.
\newblock Dynamic tolerance region computing for multimedia.
\newblock {\em IEEE Trans. Comput.}, 2012.

\bibitem{radeon}
AMD.
\newblock {Radeon GPUs}.
\newblock {\em
  http://www.amd.com/us/products/desktop/graphics/amd-radeon-hd-6000/Pages/amd-radeon-hd-6000.aspx}.

\bibitem{sc2}
A.~Arelakis et~al.
\newblock {SC2: A Statistical Compression Cache Scheme}.
\newblock In {\em ISCA}, 2014.

\bibitem{Arnau}
J.~Arnau et~al.
\newblock {Boosting mobile GPU performance with a decoupled access/execute
  fragment processor}.
\newblock In {\em ISCA}, 2012.

\bibitem{Arnau-memo}
J.~Arnau et~al.
\newblock {Eliminating Redundant Fragment Shader Executions on a Mobile {GPU}
  via Hardware Memoization}.
\newblock In {\em ISCA}, 2014.

\bibitem{medic}
R.~Ausavarangnirun et~al.
\newblock {Exploiting Inter-Warp Heterogeneity to Improve GPGPU Performance}.
\newblock In {\em PACT}, 2014.

\bibitem{stride1}
J.~Baer et~al.
\newblock Effective hardware-based data prefetching for high-performance
  processors.
\newblock {\em IEEE Trans. Comput.}, 1995.

\bibitem{GPGPUSim}
A.~Bakhoda et~al.
\newblock {Analyzing CUDA Workloads Using a Detailed GPU Simulator}.
\newblock In {\em ISPASS}, 2009.

\bibitem{gpgpu-sim}
A.~Bakhoda, G.L. Yuan, W.W.L. Fung, H.~Wong, and T.M. Aamodt.
\newblock {Analyzing CUDA Workloads Using a Detailed GPU Simulator}.
\newblock In {\em ISPASS}, 2009.

\bibitem{cudadma}
M.~Bauer et~al.
\newblock {CudaDMA: Optimizing GPU} memory bandwidth via warp specialization.
\newblock In {\em SC}, 2011.

\bibitem{ht2}
J.~A. Brown et~al.
\newblock Speculative precomputation on chip multiprocessors.
\newblock In {\em MTEAC}, 2001.

\bibitem{lonestar}
M.~Burtscher et~al.
\newblock A quantitative study of irregular programs on gpus.
\newblock In {\em IISWC}, 2012.

\bibitem{sfu2}
D.~De Caro et~al.
\newblock High-performance special function unit for programmable 3-d graphics
  processors.
\newblock {\em Trans. Cir. Sys. Part I}, 2009.

\bibitem{ssmt}
R.~S. Chappell et~al.
\newblock Simultaneous subordinate microthreading ({SSMT}).
\newblock In {\em ISCA}, 1999.

\bibitem{ssmt2}
R.~S. Chappell et~al.
\newblock Microarchitectural support for precomputation microthreads.
\newblock In {\em MICRO}, 2002.

\bibitem{rodinia}
S.~Che et~al.
\newblock {Rodinia: A Benchmark Suite for Heterogeneous Computing}.
\newblock In {\em IISWC}, 2009.

\bibitem{c-pack}
X.~Chen et~al.
\newblock C-pack: A high-performance microprocessor cache compression
  algorithm.
\newblock In {\em IEEE Trans. on VLSI Systems}, 2010.

\bibitem{memo-inst}
D.~Citron et~al.
\newblock Accelerating multi-media processing by implementing memoing in
  multiplication and division units.
\newblock In {\em ASPLOS}, 1998.

\bibitem{ht4}
J.~D. Collins et~al.
\newblock Dynamic speculative precomputation.
\newblock In {\em MICRO}, 2001.

\bibitem{ht5}
J.~D. Collins et~al.
\newblock {Speculative Precomputation: Long-range Prefetching of Delinquent
  Loads}.
\newblock {\em ISCA}, 2001.

\bibitem{danconnors}
D.~A. Connors et~al.
\newblock Compiler-directed dynamic computation reuse: rationale and initial
  results.
\newblock In {\em MICRO}, 1999.

\bibitem{noc-comp}
R.~Das et~al.
\newblock {Performance and Power Optimization through Data Compression in
  Network-on-Chip Architectures}.
\newblock In {\em HPCA}, 2008.

\bibitem{assisted-execution-04}
M.~Dubois.
\newblock Fighting the memory wall with assisted execution.
\newblock In {\em CF}, 2004.

\bibitem{assisted-execution}
M.~Dubois et~al.
\newblock Assisted execution.
\newblock Technical report, USC, 1998.

\bibitem{zca}
J.~Dusser et~al.
\newblock Zero-content augmented caches.
\newblock In {\em ICS}, 2009.

\bibitem{ebrahimi-micro09}
E.~Ebrahimi et~al.
\newblock {Coordinated Control of Multiple Prefetchers in Multi-core Systems}.
\newblock In {\em MICRO}, 2009.

\bibitem{ebrahimi-hpca09}
E.~Ebrahimi et~al.
\newblock {Techniques for Bandwidth-efficient Prefetching of Linked Data
  Structures in Hybrid Prefetching Systems}.
\newblock In {\em HPCA}, 2009.

\bibitem{ebrahimi-isca11}
E.~Ebrahimi et~al.
\newblock Prefetch-aware shared resource management for multi-core systems.
\newblock {\em ISCA}, 2011.

\bibitem{MMCompression}
M.~Ekman et~al.
\newblock {A Robust Main-Memory Compression Scheme}.
\newblock In {\em ISCA-32}, 2005.

\bibitem{stride2}
J.~W.~C. Fu et~al.
\newblock Stride directed prefetching in scalar processors.
\newblock In {\em MICRO}, 1992.

\bibitem{compiler-register}
M.~Gebhart et~al.
\newblock A compile-time managed multi-level register file hierarchy.
\newblock In {\em MICRO}, 2011.

\bibitem{energy-register}
M.~Gebhart et~al.
\newblock {Energy-efficient Mechanisms for Managing Thread Context in
  Throughput Processors}.
\newblock In {\em ISCA}, 2011.

\bibitem{unified-register}
M.~Gebhart et~al.
\newblock Unifying primary cache, scratch, and register file memories in a
  throughput processor.
\newblock In {\em MICRO}, 2012.

\bibitem{smart-refresh}
M.~Ghosh et~al.
\newblock Smart refresh: An enhanced memory controller design for reducing
  energy in conventional and 3d die-stacked drams.
\newblock MICRO, 2007.

\bibitem{manual}
{GPGPU-Sim v3.2.1}.
\newblock {GPGPU-Sim Manual}.

\bibitem{mars}
B.~He et~al.
\newblock {Mars: A MapReduce Framework on Graphics Processors}.
\newblock In {\em PACT}, 2008.

\bibitem{block-reuse}
J.~Huang et~al.
\newblock Exploiting basic block value locality with block reuse.
\newblock In {\em HPCA}, 1999.

\bibitem{gddr3}
{Hynix.}
\newblock {512M (16mx32) GDDR3 SDRAM hy5rs123235fp}.

\bibitem{GDDR5}
{Hynix.}
\newblock {Hynix GDDR5 SGRAM Part H5GQ1H24AFR Revision 1.0}.

\bibitem{ht7}
K.~Z. Ibrahim et~al.
\newblock Slipstream execution mode for cmp-based multiprocessors.
\newblock In {\em HPCA}, 2003.

\bibitem{zvc}
M.~Islam et~al.
\newblock {Zero-Value Caches: Cancelling Loads that Return Zero}.
\newblock In {\em PACT}, 2009.

\bibitem{ITRS}
{ITRS}.
\newblock International technology roadmap for semiconductors.
\newblock 2011.

\bibitem{osp-isca13}
A.~Jog et~al.
\newblock {Orchestrated Scheduling and Prefetching for GPGPUs}.
\newblock In {\em ISCA}, 2013.

\bibitem{owl-asplos13}
A.~Jog et~al.
\newblock {OWL: Cooperative Thread Array Aware Scheduling Techniques for
  Improving GPGPU Performance}.
\newblock In {\em ASPLOS}, 2013.

\bibitem{patterson}
{John L. Hennessey and David A. Patterson}.
\newblock {\em {Computer Architecture, A Quantitaive Approach}}.
\newblock {Morgan Kaufmann}, 2010.

\bibitem{stream1}
N.~Jouppi.
\newblock Improving direct-mapped cache performance by the addition of a small
  fully-associative cache and prefetch buffers.
\newblock In {\em ISCA}, 1990.

\bibitem{ht0}
M.~Kamruzzaman et~al.
\newblock {Inter-core Prefetching for Multicore Processors Using Migrating
  Helper Threads}.
\newblock In {\em ASPLOS}, 2011.

\bibitem{nmnl-pact13}
O.~Kayiran et~al.
\newblock {Neither More Nor Less: Optimizing Thread-level Parallelism for
  GPGPUs}.
\newblock In {\em PACT}, 2013.

\bibitem{kayiran-micro14}
O.~Kayiran et~al.
\newblock {Managing GPU Concurrency in Heterogeneous Architectures}.
\newblock In {\em MICRO}, 2014.

\bibitem{keckler}
S.~W. Keckler et~al.
\newblock {GPUs} and the future of parallel computing.
\newblock {\em IEEE Micro}, 2011.

\bibitem{ht11}
D.~Kim et~al.
\newblock {Design and Evaluation of Compiler Algorithms for Pre-execution}.
\newblock In {\em ASPLOS}, 2002.

\bibitem{ht10}
Dongkeun Kim et~al.
\newblock Physical experimentation with prefetching helper threads on intel's
  hyper-threaded processors.
\newblock In {\em CGO}, 2004.

\bibitem{wen-mei-hwu}
David~B. Kirk and W.~Hwu.
\newblock {\em Programming massively parallel processors: a hands-on approach}.
\newblock Morgan Kaufmann, 2010.

\bibitem{ht15}
Venkata Krishnan and Josep Torrellas.
\newblock A chip-multiprocessor architecture with speculative multithreading.
\newblock {\em IEEE Trans. Comput.}, 1999.

\bibitem{spareregister}
N.~Lakshminarayana et~al.
\newblock {Spare register aware prefetching for graph algorithms on GPUs}.
\newblock In {\em HPCA}, 2014.

\bibitem{kim}
J.~Lee et~al.
\newblock {Many-Thread Aware Prefetching Mechanisms for GPGPU Applications}.
\newblock In {\em {MICRO}}, 2010.

\bibitem{gpuwattch}
J.~Leng et~al.
\newblock {{GPUWattch}: Enabling Energy Optimizations in GPGPUs}.
\newblock In {\em ISCA}, 2013.

\bibitem{sfu}
E.~Lindholm et~al.
\newblock Nvidia tesla: A unified graphics and computing architecture.
\newblock {\em IEEE Micro}, 2008.

\bibitem{raidr}
J.~Liu et~al.
\newblock Raidr: Retention-aware intelligent dram refresh.
\newblock ISCA, 2012.

\bibitem{ht25}
J.~Lu et~al.
\newblock {Dynamic Helper Threaded Prefetching on the Sun UltraSPARC CMP
  Processor}.
\newblock In {\em MICRO}, 2005.

\bibitem{ht12}
C.~Luk.
\newblock Tolerating memory latency through software-controlled pre-execution
  in simultaneous multithreading processors.
\newblock In {\em ISCA}, 2001.

\bibitem{pin}
C.~Luk et~al.
\newblock {Pin: Building Customized Program Analysis Tools with Dynamic
  Instrumentation}.
\newblock In {\em PLDI}, 2005.

\bibitem{luo-dsn14}
Y.~Luo et~al.
\newblock {Characterizing Application Memory Error Vulnerability to Optimize
  Data Center Cost via Heterogeneous-Reliability Memory}.
\newblock DSN, 2014.

\bibitem{ht16}
Pedro Marcuello et~al.
\newblock Speculative multithreaded processors.
\newblock In {\em ICS}, 1998.

\bibitem{Meng}
J.~Meng et~al.
\newblock Dynamic warp subdivision for integrated branch and memory divergence
  tolerance.
\newblock In {\em ISCA}, 2010.

\bibitem{meza-cal}
J.~Meza et~al.
\newblock Enabling efficient and scalable hybrid memories.
\newblock IEEE CAL, 2012.

\bibitem{slice}
A.~Moshovos et~al.
\newblock Slice-processors: An implementation of operation-based prediction.
\newblock In {\em ICS}, 2001.

\bibitem{rmt-alternatives}
S.~Mukherjee et~al.
\newblock Detailed design and evaluation of redundant multithreading
  alternatives.
\newblock {\em ISCA}.

\bibitem{runahead-hpca03}
O.~Mutlu et~al.
\newblock Runahead execution: An alternative to very large instruction windows
  for out-of-order processors.
\newblock In {\em HPCA}, 2003.

\bibitem{runahead-isca05}
O.~Mutlu et~al.
\newblock Techniques for efficient processing in runahead execution engines.
\newblock ISCA, 2005.

\bibitem{veynu}
V.~Narasiman et~al.
\newblock Improving {GPU} performance via large warps and two-level warp
  scheduling.
\newblock In {\em MICRO}, 2011.

\bibitem{valgrind}
N.~Nethercote et~al.
\newblock {Valgrind: A Framework for Heavyweight Dynamic Binary
  Instrumentation}.
\newblock In {\em PLDI}, 2007.

\bibitem{coal1}
B.~S. Nordquist et~al.
\newblock Apparatus, system, and method for coalescing parallel memory
  requests, 2009.
\newblock {US Patent 7,492,368}.

\bibitem{pguide}
NVIDIA.
\newblock {Programming Guide}.

\bibitem{sdk}
NVIDIA.
\newblock {CUDA C/C++ SDK Code Samples}, 2011.

\bibitem{fermi}
NVIDIA.
\newblock {Fermi: NVIDIA's Next Generation CUDA Compute Architecture}, 2011.

\bibitem{coal2}
L.~Nyland et~al.
\newblock Systems and methods for coalescing memory accesses of parallel
  threads, 2011.
\newblock {US Patent 8,086,806}.

\bibitem{stream2}
S.~Palacharla et~al.
\newblock Evaluating stream buffers as a secondary cache replacement.
\newblock In {\em ISCA}, 1994.

\bibitem{bdi}
G.~Pekhimenko et~al.
\newblock {Base-Delta-Immediate Compression: Practical Data Compression for
  On-Chip Caches}.
\newblock In {\em PACT}, 2012.

\bibitem{lcp-micro}
G.~Pekhimenko et~al.
\newblock {Linearly Compressed Pages: A Low Complexity, Low Latency Main Memory
  Compression Framework}.
\newblock In {\em MICRO}, 2013.

\bibitem{camp}
G.~Pekhimenko et~al.
\newblock {Exploiting Compressed Block Size as an Indicator of Future Reuse}.
\newblock In {\em HPCA}, 2015.

\bibitem{toggle-aware-cal}
G.~Pekhimenko et~al.
\newblock {Toggle-Aware Compression for GPUs}.
\newblock In {\em IEEE CAL}, 2015.

\bibitem{toggle-aware-hpca}
G.~Pekhimenko et~al.
\newblock {A Case for Toggle-Aware Compression in GPUs}.
\newblock In {\em HPCA}, 2016.

\bibitem{ht17}
Carlos~Garc\'{\i}a Qui\~{n}ones et~al.
\newblock {Mitosis Compiler: An Infrastructure for Speculative Threading Based
  on Pre-computation Slices}.
\newblock In {\em PLDI}, 2005.

\bibitem{Qureshi-rmt}
M.~Qureshi et~al.
\newblock Microarchitecture-based introspection: A technique for
  transient-fault tolerance in microprocessors.
\newblock DSN, 2005.

\bibitem{qureshi-dram-caches}
M.~Qureshi et~al.
\newblock Fundamental latency trade-off in architecting dram caches:
  Outperforming impractical sram-tags with a simple and practical design.
\newblock MICRO, 2012.

\bibitem{rhu-micro13}
Minsoo Rhu, Michael Sullivan, Jingwen Leng, and Mattan Erez.
\newblock {A Locality-Aware Memory Hierarchy for Energy-Efficient GPU
  Architectures}.
\newblock In {\em MICRO}, 2013.

\bibitem{tor-micro12}
T.~G. Rogers et~al.
\newblock {Cache-Conscious Wavefront Scheduling}.
\newblock In {\em MICRO}, 2012.

\bibitem{ddmt}
A.~Roth et~al.
\newblock Speculative data-driven multithreading.
\newblock In {\em HPCA}, 2001.

\bibitem{sage}
M.~Samadi et~al.
\newblock Sage: Self-tuning approximation for graphics engines.
\newblock In {\em MICRO}, 2013.

\bibitem{dcc}
S.~Sardashti et~al.
\newblock {Decoupled Compressed Cache: Exploiting Spatial Locality for
  Energy-optimized Compressed Caching}.
\newblock In {\em MICRO}, 2013.

\bibitem{GPUBandwidthCompression}
V.~Sathish et~al.
\newblock {Lossless and Lossy Memory I/O Link Compression for Improving
  Performance of GPGPU Workloads}.
\newblock In {\em PACT}, 2012.

\bibitem{page-overlays}
V.~Seshadri et~al.
\newblock Page overlays: An enhanced virtual memory framework to enable
  fine-grained memory management.
\newblock ISCA, 2015.

\bibitem{Apogee}
A.~Sethia et~al.
\newblock Apogee: adaptive prefetching on gpus for energy efficiency.
\newblock In {\em PACT}, 2013.

\bibitem{equalizer}
A.~Sethia et~al.
\newblock Equalizer: Dynamic tuning of gpu resources for efficient execution.
\newblock In {\em MICRO}, 2014.

\bibitem{memzip}
A.~Shafiee et~al.
\newblock {MemZip: Exploring Unconventional Benefits from Memory Compression}.
\newblock In {\em HPCA}, 2014.

\bibitem{burtonsmith}
B.~Smith.
\newblock {A pipelined, shared resource MIMD computer}.
\newblock {\em Advance Computer Architecture}, 1986.

\bibitem{dynreuse}
A.~Sodani et~al.
\newblock {Dynamic Instruction Reuse}.
\newblock In {\em ISCA}, 1997.

\bibitem{ht18}
Sohi et~al.
\newblock {Multiscalar Processors}.
\newblock In {\em ISCA}, 1995.

\bibitem{BPKI}
S.~Srinath et~al.
\newblock {Feedback Directed Prefetching: Improving the Performance and
  Bandwidth-Efficiency of Hardware Prefetchers}.
\newblock In {\em HPCA}, 2007.

\bibitem{fdp}
S.~Srinath et~al.
\newblock {Feedback Directed Prefetching: Improving the Performance and
  Bandwidth-Efficiency of Hardware Prefetchers}.
\newblock In {\em HPCA}, 2007.

\bibitem{gpuPin}
M.~Stephenson et~al.
\newblock Flexible software profiling of {GPU} architectures.
\newblock In {\em ISCA}, 2015.

\bibitem{ht9}
K.~Sundaramoorthy et~al.
\newblock {Slipstream Processors: Improving Both Performance and Fault
  Tolerance}.
\newblock In {\em ASPLOS}, 2000.

\bibitem{cdc6600}
J.~E. Thornton.
\newblock {Parallel Operation in the Control Data 6600}.
\newblock {\em Proceedings of the AFIPS FJCC}, 1964.

\bibitem{cacti}
S.~Thoziyoor et~al.
\newblock {CACTI} 5.1.
\newblock Technical Report HPL-2008-20, {HP Laboratories}, 2008.

\bibitem{LinkCompression}
M.~Thuresson et~al.
\newblock {Memory-Link Compression Schemes: A Value Locality Perspective}.
\newblock {\em IEEE Trans. Comput.}, 2008.

\bibitem{caba}
N.~Vijaykumar et~al.
\newblock {A Case for Core-Assisted Bottleneck Acceleration in GPUs: Enabling
  Flexible Data Compression with Assist Warps}.
\newblock In {\em ISCA}, 2015.

\bibitem{gpu-rmt}
J.~Wadden et~al.
\newblock Real-world design and evaluation of compiler-managed gpu redundant
  multithreading.
\newblock ISCA '14, 2014.

\bibitem{wang-rmt}
Wang et~al.
\newblock Compiler-managed software-based redundant multi-threading for
  transient fault detection.
\newblock CGO, 2007.

\bibitem{fvc}
J.~Yang et~al.
\newblock {Frequent Value Compression in Data Caches}.
\newblock In {\em MICRO}, 2000.

\bibitem{RFVP}
A.~Yazdanbakhsh et~al.
\newblock Mitigating the memory bottleneck with approximate load value
  prediction.
\newblock IEEE Date and Test, 2016.

\bibitem{ht19}
W.~Zhang et~al.
\newblock Accelerating and adapting precomputation threads for effcient
  prefetching.
\newblock In {\em HPCA}, 2007.

\bibitem{zilles-exception}
C.~Zilles et~al.
\newblock The use of multithreading for exception handling.
\newblock MICRO, 1999.

\bibitem{ht20}
C.~Zilles et~al.
\newblock {Execution-based Prediction Using Speculative Slices}.
\newblock In {\em ISCA}, 2001.

\end{thebibliography}
\end{footnotesize}
\end{spacing}

\end{document}